\documentclass{article}

\usepackage[utf8]{inputenc}
\usepackage{natbib}
\usepackage{geometry}
\usepackage{graphicx}
\usepackage{subfigure}
\newcommand{\hMpc}{\ h^{-1}\mathrm{Mpc}}
\newcommand{\hMsun}{\ h^{-1}\mathrm{M}_{\odot}}

\title{Report from the Tri-Agency Cosmological Simulation Task Force}
\date{\vspace{-5ex}}

\begin{document}
\maketitle

\vspace{-1cm}

{\bf\em Authors: }{Nick Battaglia\unskip$^{1}$, Andrew Benson$^{2}$, Tim Eifler$^{3}$, Andrew Hearin$^{4}$, Katrin Heitmann$^{4}$, Shirley Ho$^{5,11,12}$, Alina Kiessling$^{6}$, Zarija Luki\'c$^{7}$, Michael Schneider$^{8}$, Elena Sellentin$^{9}$, Joachim Stadel$^{10}$ 
}
\vspace{0.4cm}\\
\small{
$^{1}$ Cornell University, Ithaca, NY 14853, USA\\
$^{2}$ Carnegie Observatories, Pasadena, CA 91101, USA\\
$^{3}$ Steward Observatory, University of Arizona, Tucson, AZ 85721, USA\\
$^{4}$ Argonne National Laboratory, Lemont, IL 60439, USA\\
$^{5}$ Flatiron Institute, New York, NY 10010, USA\\
$^{6}$ Jet Propulsion Laboratory/California Institute of Technology, Passadena, CA 91009, USA\\
$^{7}$ Lawrence Berkeley National Laboratory, Berkeley, CA 94720, USA\\
$^{8}$ Lawrence Livermore National Laboratory, Livermore, CA 94550, USA\\
$^{9}$ Leiden Observatory, Leiden NL-2333, The Netherlands\\
$^{10}$ University of Zurich, Zurich 8057, Switzerland\\
$^{11}$ Princeton University, Princeton, NJ 08540, USA \\
$^{12}$ Carnegie Mellon University, Pittsburgh, PA 15213, USA \\

}

\begin{center}
    {\bf Foreword}
\end{center}
The Tri-Agency Cosmological Simulations (TACS) Task Force was formed when Program Managers from the Department of Energy (DOE), the National Aeronautics and Space Administration (NASA), and the National Science Foundation (NSF) expressed an interest in receiving input into the cosmological simulations landscape related to the upcoming DOE/NSF Vera Rubin Observatory (Rubin), NASA/ESA's Euclid, and NASA's Wide Field Infrared Survey Telescope (WFIRST). The Co-Chairs of TACS, Katrin Heitmann and Alina Kiessling, invited community scientists from the USA and Europe who are each subject matter experts and are also members of one or more of the surveys to contribute. The following report represents the input from TACS that was delivered to the Agencies in December 2018.

\begin{center}
{\bf Executive Summary}
\end{center}
Upcoming wide-field surveys require extensive numerical simulations for a number of interrelated tasks, including carrying out the simulations, transforming them into synthetic sky maps, validating the results, and serving the data in an easily accessible way. These are all major efforts involving large computing and storage resources as well as people with specialized expertise to develop the modeling and analysis pipelines and database approaches. Many of the tasks are common between the major cosmological surveys
and it is therefore strongly advisable to evaluate common approaches and resource sharing between the surveys. Additionally, investigations of scientific gains that can be reaped from joint pixel analysis efforts have been initiated; such investigations rely on the availability of shared synthetic catalogs that can be used across the surveys and are based on the same underlying cosmological simulations.

Here we report on our findings regarding common generation, use, and curation of cosmological simulation data products for the Vera Rubin Observatory (Rubin) Dark Energy Science Collaboration (DESC), the Wide Field Infrared Survey Telescope (WFIRST), and Euclid, as well as possibilities to share in a common computational and storage infrastructure. We describe the use of extreme-scale simulations (defined as simulations that require very large computational resources)
as well as simulation suites that enable covariance estimation and exploration of the cosmological and physical modeling parameter space. We then discuss different methods for generating synthetic sky maps from gravity-only simulations. While some aspects of these methods must be tailored to the survey at hand, the general concepts that are being developed are similar and sharing resources to develop those concepts would be natural. Next we describe the biggest challenge facing cosmology in the coming decade -- understanding systematic effects and disentangling them from cosmological information. In this report we focus on systematic effects that can be addressed by simulation efforts and are common across the three surveys.
Another area where joint efforts can be fruitful is in the exploration of statistical methods. Here we discuss a range of topics from next generation prediction tools to covariances. Finally, we describe the advantages of a common infrastructure to share simulation products.  

Our report should make it clear that providing joint resources between the surveys will enable efficient development and sharing of simulations and related analysis tools. The current support for a program of this nature is not well established since often these activities are viewed as infrastructure tasks rather than as a broader research and development activity. Consequently, funding that, in particular, supports work across surveys (and therefore funding Agencies) is sparse. However, we also emphasize that survey-specific work still needs to be supported as well. In order to make this distinction between cross-survey and survey-specific work very clear,  each subsection of our report ends with a summary of our findings with particular focus on how cross-survey support would strengthen the specific research and development described.

\newpage
\tableofcontents
\newpage

\section{Motivation}

Cosmological simulations have become increasingly sophisticated over the last several decades and their role in cosmological surveys has correspondingly experienced enormous growth. Cosmological simulations are now integral to forecasting and survey formulation, in addition to the eventual analysis of the observational data. The shift from Stage 3 to Stage 4 cosmological surveys has been underway for the last several years and during this time the role of cosmological simulations in the surveys has undergone a shift from being a research and development (R\&D) effort to being a key element of the Stage 4 survey infrastructure. 

Elements that are considered part of the survey infrastructure are deemed as essential to the success of the survey and have traditionally included efforts like ground operations, analysis pipelines, and data management pipeline development, but not cosmological simulations. However, although it is currently widely accepted that cosmological simulations are essential to upcoming Stage 4 surveys, the funding and support for these efforts is still largely only being covered by competitively selected R\&D proposals. As a result, key work is difficult to undertake in a timely or planned manner due to the uncertainty of proposal selection. This has resulted in efforts to date being limited to the few groups that have been successful in securing short-term funding and resources for very specific tasks. 

Added to this challenge is the reality that students and postdocs working on cosmological simulations and synthetic sky generation have historically had very little success in securing permanent jobs in the field. Consequently, the number of people available to contribute to these efforts is consistently low and the ``next generation'' are being lost to more secure jobs in data science. While we do not attempt to solve this issue directly in this document, we did want to highlight it as a pervasive problem that is deserving of more focused consideration in future.  

The purpose of this document is to clearly detail cosmological simulation efforts that are essential to the success of the upcoming Stage 4 cosmological surveys from the Vera Rubin Observatory (Rubin), the Wide Field Infrared Survey Telescope (WFIRST), and Euclid. The document also highlights work that is still required and focuses on collaborative efforts that will benefit two or more of the surveys. A focused collaboration between the surveys and Agencies will enable the most efficient use of resources and will facilitate rapid development in key areas that are currently experiencing only moderate progress due to a lack of support. It is important to stress that such a collaboration does need additional support that is currently not available. 

The document begins by introducing ``extreme-scale'' simulations followed by large simulation campaigns, which are the two primary classes of simulation required for upcoming cosmological surveys. Next, the generation of synthetic sky maps and the challenges to this effort are discussed in detail, followed by an analysis of how simulations are essential to investigating and mitigating systematic effects. The role of simulations in developing advanced statistical techniques is then investigated and the document concludes by presenting an argument for the development of a common infrastructure to share simulation products.  

\section{Extreme-scale Simulations}\label{sec:hero}

\subsection{Introduction}
In this section we discuss what we call ``extreme-scale simulations'' -- very large, high-resolution N-body simulations  (``grand challenge" simulations) that form the basis for synthetic sky maps and very detailed, large hydrodynamic simulations that are important to advance our understanding of astrophysics systematics. These simulations require major computing allocations (in the U.S. for example, the DOE INCITE -- Innovative and Novel Computational Impact on Theory and Experiment -- Program provides opportunities to apply for such allocations at the Leadership Computing Facilities) in the tens of millions of hours (exact numbers depend on the supercomputer) and access to a supercomputer with a performance that is close to the top 10 supercomputers in the world\footnote{https://www.top500.org/list/2018/06/?page=1}.
By the very definition of extreme-scale, only a handful of these simulations will be available in the world at any one time due to the high cost of running the simulations and storing the outputs. We first discuss gravity-only simulations and focus on questions concerning the construction of synthetic catalogs for cosmological surveys. The details on how the catalogs are built are given in Section~\ref{sec:maps} and some of the requirements listed here are justified in that section in more detail. Next we discuss hydrodynamic simulations for cosmological surveys. This area still requires significantly more development to ensure that the physics  in the simulations is captured  correctly and to enable simulations of cosmological volumes at sufficient resolution. Detailed use-cases of the hydrodynamic simulations in the context of cosmological surveys are discussed in Section~\ref{sec:syst}. 

\subsection{Gravity-only Simulations}\label{sec:GO}
We start by summarizing the requirements from the three surveys, focusing on volume and mass resolution for gravity-only simulations. Next, we list the outputs obtained after an initial analysis step to enable the investigation of the different cosmological probes targeted by the surveys. In this case, the requirements are essentially the same for all surveys -- if there are specific needs for a subset of the surveys, these will be explicitly listed. Currently, two codes are actively being developed within the cosmology community that can carry out these ``base'' simulations at the needed resolution, PKDGRAV3 described in~\cite{potter} and HACC (Hardware/Hybrid Accelerated Cosmology Code) described in~\cite{habib}. While there are obviously many more simulation codes, most of them do not scale to the largest supercomputers currently available (scaling is a prerequisite when applying for large supercomputer allocations at programs such as the INCITE Program) or cannot take advantage of the architectures at all (e.g. only very few N-body codes can currently run on GPU-accelerated systems)\footnote{2HOT is another code that has GPU capabilities but the code author left the field and the code is currently not being actively developed.}. In addition, in order to carry out these large simulations, the memory footprint of the codes has to be optimized to enable the simulation of trillions and more particles. The development of these kinds of scalable codes, including analysis tools that can be run on the fly to minimize storage requirements, usually takes major development efforts that continue over many years to adjust to new computing hardware. Both PKDGRAV3 and HACC can take full advantage of current accelerated supercomputing architectures. The remainder of this section describes the already available simulations and possible future simulations, including plans going beyond $\Lambda$CDM. 

\subsubsection{Requirements and Outputs}\label{sec:hero:req}
The exact requirements and outputs needed to generate synthetic sky maps depends strongly on the method used to populate the simulation with the objects of interest, as described in Section~\ref{sec:maps}, and the targeted wavelength. In this report we focus on optical and near-IR surveys, even though for cross-correlation tasks, other wavelengths are an important target for synthetic sky maps as well. With the type of simulations described in this section, galaxy modeling approaches focus on using the galaxy-halo connection rather than methods that identify galaxies with single particles. Given the depth of the surveys of interest and the desire to resolve halos with a minimum number of particles to host a galaxy, the simulation requirements are rather demanding. 

First, we briefly describe the requirements with regard to volume and mass resolution for the simulations that underlie detailed synthetic sky catalogs as required by Rubin, Euclid, and WFIRST. All three surveys are exceptionally deep and/or wide and therefore require large volume simulations to capture the complete survey volume out to the desired redshifts. At the same time, the resolution of the dim galaxies that will be captured in the surveys requires the simulation to have a very high particle mass resolution. The volume for these simulations should not be less than 3~Gpc to avoid too many replications when creating light cones at high redshifts. Optimally, the volume covered would be around 5~Gpc, which is difficult to reach with currently available supercomputers at the required mass resolution. As we discuss further below, most methods for generating synthetic catalogs rely on the identification of subhalos and on finding halos down to low masses. Therefore the (minimal) particle mass resolution requirement for an extreme-scale simulation is around $10^9$M$_\odot$, more optimal would be $10^8$M$_\odot$, which is, however, difficult to reach in large volume simulations. Recently developed methods to generate synthetic catalogs for Rubin-DESC (Dark Energy Science Collaboration) employ a hybrid approach that uses results from smaller volume simulations with mass resolutions around $10^8$M$_\odot$ to populate halo catalogs with less well resolved halos. 

Next, we describe the data products that are extracted from the N-body simulations to prepare the creation of synthetic catalogs. These catalogs include galaxy properties as well as weak lensing shear measurements. In order of increasing complexity, the products typically include:
\begin{enumerate}
\item \emph{Particle lightcone data:} These are required for the generation of shear maps. In principle, they can be constructed after the simulations finish if enough particle snapshots are saved. In practice, the amount of data that would need to be stored from these large runs is in the few PBytes which is very challenging for most computing centers. Therefore, it is preferable to generate the lightcones on the fly. We discuss some more technical questions about this below.
\item \emph{Dark matter halo positions and masses:} These are required by all synthetic sky modeling approaches considered here (with the exception of hydrodynamic simulations). A variety of methods have been developed to identify and characterize halos in N-body simulations. Beyond differences in halo mass arising from the somewhat arbitrary definition of a halo, properties of halos (e.g. mass, position, maximum velocity of the rotation curve) are robustly determined by almost all halo finders (see \cite{2011MNRAS.415.2293K} for an extensive halo finder comparison), except under conditions of major mergers \citep{2015MNRAS.454.3020B} where careful consideration of the algorithm used is required. Given the need to push base simulations to the limits of current supercomputing facility's abilities to achieve survey simulation goals, there remains a fundamental limit to the accuracy with which halo properties can be recovered due to finite particle sampling \citep{trenti_how_2010,2017MNRAS.467.3454B}.
\item \emph{Dark matter subhalo positions and masses:} Subhalo masses are not required by classic halo occupation distribution (HOD) approaches, but are required by both subhalo abundance matching (SHAM) and semi-analytic model (SAM)\footnote{This is true for most SAMs, although some are able to operate without information on subhalos, in which case they either provide no information on the spatial distribution of galaxies within each halo (beyond simply identifying one galaxy as the central), or determine this information by integrating subhalo orbits directly.} approaches. As with halos, a variety of tools have been developed to identify and quantify subhalos in N-body simulations. In general, subhalo masses are determined robustly down to around 100 particles, with subhalo detection robust down to as few as 20 particles \citep{2012MNRAS.423.1200O}. Other important quantities, such as subhalo spins \citep{2013MNRAS.429.2739O} are determined robustly, while the spatial distribution of subhalos display discrepancies at the 5--10\% level between different finder algorithms \citep{2014MNRAS.438.3205P}.
\item \emph{Merger trees:} Providing the linkage between halos across time, merger trees are required for all SAMs, and by some empirical models. Construction of merger trees is non-trivial and requires careful consideration of how to identify progenitor/descendant halos across multiple snapshots of the base simulation \citep{2013MNRAS.436..150S,2016MNRAS.459.1554W}, and relies crucially on the properties of the input halo/subhalo catalogs \citep{2014MNRAS.441.3488A}.
\item \emph{Halo shapes:} While halo shapes (i.e. the departure of the halo from spherical symmetry) are not directly required by any SAM or empirical model that we know of, they are often used to assign position angles to galaxies---an important consideration for weak lensing studies which must assess the importance of intrinsic alignments \citep{2015SSRv..193...67K}. Shape determination is known to be affected by the number of particles with which a halo is resolved in the base simulation \citep{2012JCAP...05..030S}, but the consequences of this for synthetic sky simulations have not been assessed. Alternative methods such as using a measurement of the local tidal field smoothed on scales of ~300 kpc or larger are currently being developed and would provide an approach that relies on the simulation raw output particle data instead of halo information.  
\end{enumerate}

The details of the merger tree construction algorithm can have significant effects on the properties of the resulting galaxies \citep{2014MNRAS.445.4197L}. While these systematic effects can often be ``calibrated out'' by adjusting the parameters of the galaxy model, calibration is expensive, and limits the applicability of the model to a single combination of base simulation, halo finder, and tree builder---this may be limiting if a model is to be applied to synthetic sky catalog generation for multiple surveys.

Finally, the finite number of discrete particle snapshots of the base simulations which are typically stored can have consequences for the resulting galaxy catalogs. \cite{benson_convergence_2012} emphasize the need for a minimum number of snapshots to ensure that galaxy properties are converged (a requirement that becomes more problematic at high redshifts where fewer outputs prior to that redshift are available). The discreteness of base simulation snapshots also propagates into the construction of lightcone catalogs---in which galaxy properties are output at the epoch at which the galaxy crosses the past lightcone of an ``observer''---in both the positions, physical properties, and observable properties (e.g. SEDs) of galaxies. While these discreteness effects can be minimized through interpolation\footnote{Or, of course, by simply outputting more snapshots of the base simulation, although this will be limited by available data storage, and can lead to challenges in building merger trees \protect\citep{2016MNRAS.459.1554W}.} \citep{2013MNRAS.429..556M}, they are difficult to fully mitigate. In the Euclid Flagship simulation (see \citet{potter} and the description in Section~\ref{sec:sims}) this problem was overcome by generating a particle light cone on-the-fly (while the simulation was running) and finding halos directly using the particle lightcone data. This approach also addresses some of the storage concerns (since a large number of full raw particle snapshots does not need to be stored) though also restricts the available number of light cones (if the lightcones are generated in post-processing, the observer can be placed in many locations and therefore many light cones can be generated).

\subsubsection{Currently Available Simulations}
\label{sec:sims}
Currently, there are three major simulations available that cover large volumes at high mass resolution and are being used for generating synthetic maps for different surveys: (i) the Euclid Flagship Simulation, which is the base for the synthetic sky catalogs used by Euclid, (ii) the Outer Rim simulation, which is currently being used for the second Rubin-DESC Data Challenge and has been used for DESI and eBOSS catalogs, and (iii) the Dark Sky Simulation, which is used for building DESI catalogs.\medskip

{\bf\em The Euclid Flagship Simulation}

The Euclid Flagship Simulation (see \citealt{potter}) features a simulation box of 3780 $h^{-1}$Mpc on a side with $12,600^3$ particles, leading to a mass resolution of $2.4 \times 10^9$ $h^{-1}$M$_{\odot}$. An agreed upon reference cosmology, close to Plank 2015 values, was used with the following parameters: $\Omega_m = 0.319, \Omega_b = 0.049, \Omega_{\rm CDM} = 0.270, \Omega_{\Lambda} = 0.681, w = -1.0, h = 0.67, \sigma_8 = 0.83, n_s = 0.96$. A contribution to the energy density from relativistic species in the background was ignored ($\Omega_{{\rm RAD},\nu} = 0$). Using this Euclid Reference Cosmology allows comparison to many other smaller simulations from N-body codes as well from approximate techniques that also use these reference values within the Euclid collaboration. The initial conditions were realized at $z = 49$ with second order Lagrangian perturbation theory (2LPT) displacements from a uniform particle grid. The transfer function was generated at $z = 0$ by CAMB and the resulting power spectrum was scaled back to the starting redshift of $z = 49$ via the scale independent growth factor. The main data product was produced on-the-fly during the simulation and is a continuous full-sky particle light cone (to $z=2.3$), where each particle was output exactly when the shrinking light surface sweeps by it. This resulting ball of particles contains 10 trillion particle positions and peculiar velocities (240 TB), and it was used to compute the dark matter halos (Rockstar and FoF) and lensing maps (HealPix) that are used for generating the mock galaxy catalogs for the Euclid Consortium. Further data available includes 100 dark matter halo catalogs (that were identified using a friends-of-friends, FoF, algorithm) and power spectra at proper time snapshots, as well as 11 complete particle snapshots from $z=0.764$ to $z = 0$. This simulation was performed using PKDGRAV3 on the Piz Daint supercomputer at the Swiss National Supercomputer Center (CSCS) in 2016.\medskip

{\bf\em The Outer Rim Simulation}

The Outer Rim simulation covers a volume of (3$h^{-1}$Gpc)$^3$ and evolves 10,240$^3$ particles, leading to a mass resolution of $\sim 1.85\cdot
10^9 h^{-1}$M$_\odot$. The simulation was carried out on the Argonne Leadership Computing Facility's BlueGene/Q machine, Mira, in 2013/14. Almost 100 time snapshots were saved and analyzed, yielding a data volume of more than 5PB. The data products from the simulation include halo catalogs for different mass definitions,
subhalo catalogs, detailed merger trees, two-point statistics, light cone representations of the data (halos and particles), and subsamples of raw and halo particles. The Outer Rim run continues in the tradition of the Millennium simulation by~\cite{2005Natur.435..629S}, with a similar mass resolution but with a volume coverage increased by more than a factor of 200. This is essential for capturing galaxy clustering at large length scales and for achieving the needed statistics for cluster cosmology. The cosmology used for the simulation is close to the best-fit model determined by WMAP-7~\citep{2011ApJS..192...18K}. The chosen cosmological parameters are: $\Omega_{\rm m}=0.2648$,
$\Omega_{\rm b}=0.0448$, $n_s=0.963$, $h=0.71$, $\sigma_8=0.8$, $w=-1.0$.\medskip

{\bf\em The Dark Sky Simulation}

The Dark Sky Simulation covers a much larger volume simulation box of 8 $h^{-1}$Gpc on a side with $10,240^3$ particles, leading to a mass resolution of $3.9 \times 10^{10}$ $h^{-1}$M$_{\odot}$ (see \citealt{Skillman}). The cosmological parameters are: $\Omega_b = 0.04676,~\Omega_m = 0.29504,~ \Omega_{\Lambda} = 0.70487,~w = -1.0,~h = 0.688062,~\sigma_8 = 0.8344,~n_s = 0.9676$. Light cone particle data (to $z = 2.3$) as well as Rockstar halo catalogs are available for this simulation ($\approx$ 16 TB). While the Dark Sky Simulation evolved an impressive number of particles, its mass resolution is not sufficient for the mock galaxy catalogs needed by current and upcoming surveys. The dark sky simulation was performed using the 2HOT code \citep{Warren} on Titan at the Oak Ridge Nation Labs Supercomputer Center.

\subsubsection{Future Needs}

Both Euclid and Rubin-DESC have used the large scale simulations described above to generate detailed synthetic mocks. WFIRST is currently discussing with Rubin-DESC how to take advantage of some of this work but is overall not as far advanced with their effort. Both Euclid and Rubin-DESC have started to identify future simulation needs already, which we briefly describe below. We emphasize that these descriptions are only capturing the next steps but not the full need for simulations in the future.
\medskip

{\bf\em Euclid Consortium}

There is already a need for an improved version of the Flagship Simulation which will have a mass resolution of $10^9$ $h^{-1}$M$_{\odot}$ in the same volume to satisfy the wide (15,000 sq. degree) part of the survey. Such a simulation should now also include radiation and massive neutrinos ($\sum_{\nu} m_{\nu} = 0.06$ ev) in the background as well as a linear treatment of the evolving neutrino fluctuations (and their effect on the dark matter). Halo merger trees including so called ``orphan galaxies", which attempt to continue tracing positions of dissolved subhalos, are also needed to support more sophisticated mock galaxy catalogs using SAMs (see Section~\ref{sec:maps}). Finally, there is a need for a simulation which has an order of magnitude better mass resolution ($10^8$ $h^{-1}$M$_{\odot}$) and greater depth in the light cone ($z=5$) but only covering a much smaller area on the sky in order to satisfy the needs of the deep survey (40 sq. degrees) of Euclid. One or more pencil beams could be generated from a smaller volume simulation to cover these needs, but such a simulation should statistically ``match onto" the full sky simulation and hence should use the same reference cosmological parameters. Mock galaxy catalogs for these two requirements should be made available as early as mid 2019. Combined, these 2 simulations will require about 1.2 million node-hours (with 1 GPU per node), or 50 million core hours without GPUs. The storage requirements are dominated by the ``wide" simulation, having a minimum size of 500 terabytes but with the saving of a few 
snapshots will reach 1 petabyte.
\medskip

{\bf\em Rubin-DESC}

Rubin-DESC is starting to prepare a third data challenge (DC3). The Outer Rim simulation is being used for the second data challenge (DC2) to generate a 5000 sq degree synthetic sky catalog. DC3 is supposed to cover the full survey area (currently being planned at 18,000 sq degree). The Outer Rim simulation can be only used for this purpose if it is replicated many times which could lead to artifacts. On the other hand, increasing the volume considerably at the same mass resolution would be very challenging. In addition to a $\Lambda$CDM universe, DESC is planning to use a non-$\Lambda$CDM simulation to investigate the sensitivity of the different probes to subtle changes in cosmologies. For this, a new simulation (the Knowhere simulation) has been started on the Cori supercomputer (located at DOE's NERSC facility) with a mass resolution of approximately 5.5$\cdot 10^8 h^{-1}$M$_\odot$ and a box side length of 2$h^{-1}$~Gpc. While the simulation volume is not very large, the mass resolution is excellent and will enable the use of a diverse set of galaxy-halo connection approaches.

\subsection{Hydrodynamic Simulations}
\label{sec:hydro}

There are several large cosmological simulations that include baryons (and associated astrophysical processes) which attempt to provide a partially predictive model for galaxy formation, follow the evolution of baryons inside and out of galaxies, and produce the observable properties of galaxies across cosmic time. These simulations solve the hydrodynamic equations in addition to gravity and employ comprehensive physical ``subgrid" models for process including (but not limited to) radiative cooling, star formation, and feedback. We describe several examples of such simulations below. The simulations are extremely computationally expensive in order to resolve the necessary galactic scales, while still having a large statistically representative volume. However, cosmological  hydrodynamic simulations do not resolve the scales necessary to perform ab initio calculations of critical physical processes of galaxy evolution like star formation. Instead they include subgrid modeling schemes that attempt to capture the key features of the underlying physical mechanisms or simply use phenomenological prescriptions. Current hydrodynamical simulations have to push the boundaries of computational power to increase the dynamic range of the simulations in order to decrease the assumptions included in the subgrid modeling. As a result, we are limited in the number of cosmological hydrodynamic simulations that can feasibly be run.

Given the computational cost of running a single cosmological volume hydrodynamic simulation\footnote{For example Illustris used 19 million core hours on the CURIE supercomputer in France.}, much less a suite of simulations spanning cosmological and subgrid nuisance parameters, their important role in future wide-field surveys will be to characterize systematic uncertainties and provide critical tests for techniques to mitigate these uncertainties (see Section~\ref{sec:syst}). 

Although this is not the focus of this report, other surveys spanning the electromagnetic spectrum from X-rays to radio wavelengths will be carried out on similar time scales to the optical and near-IR surveys mentioned here. Combining these multi-wavelength surveys via cross-correlation measurements will provide powerful constraints on both cosmological and astrophysical parameters, for example cross-correlation measurements of optical and cosmic microwave background data has the potential to constrain the sum of neutrino masses or feedback from Active Galactic Nuclei \citep[e.g.,][]{Spacek2016,BFSS2017}. For such measurements, hydrodynamic simulations are essential to provide testable predictions to check the subgrid modeling assumptions within the simulations. Additionally, these cross-correlation measurements will provide lasting constraints that will provide critical test for and inform future sub-grid models.

\subsubsection{Requirements and Outputs}

Currently hydrodynamic simulations do not have the combination of mass resolution and volume to meet the eventual requirements for the systematic studies discussed in Section~\ref{sec:syst}. There are some simulations highlighted in the this report (see Table~\ref{tab:hydrolist}) that have sufficient mass resolution but lack volume and vice versa. However, the more immediate concern is that all of these simulations are only as good as the sub-grid physics models they employ. Further development and exploration of various models and techniques are essential to capture physical processes including star-formation and feedback at high accuracy. In addition to the information needed from N-body simulations, catalogs of simulated galaxies with optical and near-IR properties and thermodynamic properties are critical to providing necessary multi-wavelength, cross-correlation predictions and post-dictions with observations. 

\subsubsection{Currently Available Simulations}

Cosmological hydrodynamic simulations can broadly be split into those simulations which resolve galaxy properties, like morphology, and those that do not (hereafter we use the adjectives high-resolution and low-resolution to distinguish between these simulations, respectively). In Table~\ref{tab:hydrolist} we provide the specifications for the example simulations we discuss below.

\medskip

\begin{table*}
  \caption{List of large-scale cosmological hydrodynamical simulations and their specifications }
  \label{tab:hydrolist}
  \begin{center}
   \leavevmode
   \begin{tabular}{c|c|c|c|c|c} 
     \hline \hline
Simulation & Box Length & \# of DM & \# of Gas & DM Mass & Gas Mass \\ 
Name & [$\hMpc$] & Particles & Particles & [$\hMsun$] & [$\hMsun$]  \\ 
     \hline
     \multicolumn{6}{l}{High-resolution Hydrodynamical Simulations} \\
     \hline
BlueTides$^a$ & 400 & $7040^3$ & $7040^3$ & $1.2 \times 10^7$ & $2.33 \times 10^6$ \\ 
EAGLE & 67.77 & $1504^3$ & $1504^3$ & $6.57 \times 10^6$ & $1.23 \times 10^6$ \\
MassiveBlack-II & 100 & $1972^3$ & $1972^3$ & $1.1 \times 10^7$ & $2.2 \times 10^6$ \\
OWLS & 100 & $512^3$ & $512^3$ & $4.06 \times 10^8$ & $8.66 \times 10^7$ \\
Horizon AGN$^b$ & 100 & $1024^3$ & N/A & $8 \times 10^7$ & N/A \\
Illustris & 75 & $1820^3$ & $1820^3$ & $4.41 \times 10^6$ & $8.87 \times 10^5$ \\
MUFASA & 50 & $512^3$ & $512^3$ & $9.6 \times 10^7$ & $1.82 \times 10^7$ \\
\hline
\multicolumn{6}{l}{Low-resolution Hydrodynamical Simulations} \\
\hline
BAHAMAS & 400 & $1024^3$ & $1024^3$ & $4 \times 10^9$ & $8 \times 10^8$ \\
Magneticum & 2688 & $4536^3$ & $4536^3$ & $1.3 \times 10^{10}$ & $2.6\times 10^9$ \\
\hline
   \end{tabular}
  \end{center}
  \begin{quote}
    \noindent 
Note this is not a complete list of all available simulations
$^a$BlueTides was run to $z=8$.
$^b$Horizon AGN was run with an AMR code and does not use gas particles, the equivalent spacial resolution is 1 kpc (proper units)
 \end{quote}
\end{table*}

{\bf\em High-resolution hydrodynamical simulations}

The general goal of high-resolution cosmological hydrodynamical simulations is to provide a predictive model for galaxy formation and produce the detailed properties of galaxies we observe. There are several ongoing efforts to this end and the following are examples of such efforts. 

The BlueTides simulation \citep{Feng2016} aims to simulate the first galaxies and active galactic nuclei (AGN), and their contribution to reionization using a version of Lagrangian TreePM-SPH code Gadget-2 \citep{Gadget2}. This simulation is quite large given its mass resolution, however it has only been run to $z=8$ limiting its utility to high-redshifts. BlueTides was run on the Blue Waters system at the National Center for Super-computing Applications (NCSA) using the a total of 648 000 Cray XE compute cores. 

The EAGLE project \citep{Eagle} is a suite of hydrodynamical simulations that follow the formation of galaxies and supermassive black holes in cosmologically representative volumes using a version of Gadget-2. The Eagle simulation includes sub-grid physics models that are tuned to agree with key observations of galaxies properties at as close to the level possible that one could attain by semi-analytic models \citep{Eagle}. The subgrid physics used in the EAGLE simulations are based on the OWLS project \citep{Owl}, which is a large suite of simulations with varying sub-grid physics to investigate the effects of altering or adding a single physical process on the total matter distribution. The MassiveBlack-II simulation \citep{MB2} is the same size as the EAGLE simulations with slightly higher resolution. A single set of subgrid physics models was used in the MassiveBlack-II simulation.

Other projects that use different hydrodynamic solvers include the Horizon-AGN suite of simulations \citep{Horizon} which was carried out with the adaptive mesh refinement (AMR) code RAMSES \citep{RAMSES}. The Horizon-AGN simulations are similar in size and resolution compared to the other simulations described. They include a variety of subgrid models to capture baryonic processes. The Illustris simulation \citep{Illustris} used the moving mesh code AREPO \citep{Volker2010}. The Illustris simulation incorporates a broad range of galaxy formation physics \citep{Vogelsberger2013} tuned on smaller volume simulations to match stellar luminosity functions and optical properties of galaxies. The MUFASA suite of simulations \citep{Mufasa} employs the GIZMO meshless finite mass (MFM) code \citep{Hopkins2015}. Despite their size the MUFASA simulations include subgrid models that were refined on high-resolution simulations of individual galaxies from the FIRE suite of simulations \citep{Hopkins2014}. \medskip

{\bf\em Low-resolution hydrodynamical simulations}

The general goal of low-resolution hydrodynamical simulations is to follow the evolution of baryons inside and out of galaxies and to capture rare objects like clusters and super-clusters of galaxies. The lower resolution reduces the computational cost of the simulations and therefore enables simulations in larger volumes that can capture these rare objects. There are several ongoing efforts and we list some examples in the following. 

The BAHAMAS project \citep{Bahamas} is a suite of simulations run with a version Gadget-2, that have been calibrated to reproduce the present-day galaxy stellar mass function and the hot gas mass fractions of groups and clusters in order to ensure the effects of feedback on the overall matter distribution are broadly correct. The Magneticum simulations \citep[e.g.,][]{Magnet} are a suite of simulations run with a Gadget variant that have large simulation volumes with comparable resolution to the BAHAMAS simulations. The largest Magneticum simulation has a simulation box length that is roughly 6.5 times larger than a BAHAMAS box. Given the computational expense of generating this large volume simulation, only a single set of sub-grid physics models were employed. 

\subsection{TACS Findings for Extreme-scale Simulations}

\subsubsection{Gravity-only Simulations}

Gravity-only simulations at large volume and high mass resolution are extremely important for all
three surveys to enable the generation of detailed synthetic sky maps. Currently, two simulations
are available (the Euclid Flagship simulation and the Outer Rim simulation) that are used for this
purpose and are very close to the ultimately required mass resolution and volume for generating
these maps. With the advent of the next-generation supercomputers (e.g., Summit at the Oak Ridge
Leadership Computing Facility) the remaining needed increase in resolution
should be achievable relatively easily. Currently two codes are being actively developed
(PKDGRAV3 and HACC) that can carry out these extreme-scale simulations. Sharing the results
from these simulations is very desirable as these simulations are very computationally expensive to produce, analyze, and store, and there are very few people with the expertise to undertake these efforts. However, sharing will require an infrastructure support investment to enable sharing of the simulation data and to enable the collaborations to generate synthetic
catalogs given the different approaches used by the two codes to carry out analysis tasks. The
infrastructure support would include storage space accessible across the collaborations and people
support to develop an infrastructure that allows for easy data access (more details are discussed in
Section 6). Additionally, if these simulations are directly shared between the surveys (rather than
making them world-wide publicly available) the question arises of how the simulation groups should
be acknowledged for their work. At a minimum, they should be made external collaborators to
the surveys to facilitate co-authorship on papers that are enabled via their contributions.

\subsubsection{Hydrodynamic Simulations}
\label{subsubsec:hydro_assessment}

Unlike gravity-only simulations, hydrodynamic simulations are far from the ultimate goal with respect to achieving large, cosmological volume simulations at high resolution with reliable physics implementations. Not even the next-generation of supercomputers will rectify this situation, although some progress is being made to (at least) generate consistent results across codes at moderate scales. The challenges are on many fronts. First, most hydrodynamic codes do not scale efficiently to utilize the full machines available today. Given the mass resolution requirements (around 10$^7$M$_\odot$), load-balancing is a difficult task and therefore enabling large volume simulations at very high mass resolution is currently out of reach. Beyond this problem, an even more serious problem is due to the uncertainties in the current subgrid model implementations. The use of relatively crude subgrid models prevents us from achieving truly first principle predictions and therefore makes it very difficult to use the simulations for the purpose they are primarily needed for -- understanding astrophysical systematics. These systematics will ultimately be the limiting factors to improving the cosmological constraints. Therefore, it is crucial to have concerted support across the surveys for improving hydrodynamic simulation capabilities. Efforts are needed to help bridge the work carried out on the smallest scales to the larger volume, cosmologically relevant, simulations. Detailed studies of subgrid models must also be carried out to improve our understanding of baryonic effects. The most effective studies will come from multi-wavelength comparisons including cross-correlations with observables for which hydrodynamic simulations make testable predictions. Sharing the results of hydrodynamic simulations is much easier than for the gravity-only simulations due to their current limitations in size. Therefore, in order to make progress in the field of hydrodynamic simulations, emphasis should be placed on supporting code development efforts, the calibration of subgrid models, and public access to the simulations to enable wide utilization and cross-comparisons.

\subsection{Conclusions}

In this section we summarize the resource requirements for cross-survey activities for both gravity-only and hydrodynamic simulations. We emphasize that cross-survey work is currently not explicitly supported and usually only occurs if the contributing scientists belong to more than one project. The demands that each survey puts on members of the simulation team are already very high and the efforts are not supported sufficiently within each survey to begin with. Therefore, additional efforts would need to be funded to enable cross-survey collaborations.

\subsubsection{Gravity-only Simulations}

\begin{itemize}
\item \textbf{Phase 1: Definition; $\mathbf{\sim}$6 months}\\
During the first phase, the list of requirements and outputs that has been outlined in this report will need to be fleshed out to ensure that all the requirements are met for the different surveys. Definitions and units have to be agreed upon (or at least translations between different code outputs) so that the simulations can be seamlessly shared between the surveys. This requires close collaboration between the surveys and strong engagement from the working groups that will use the simulations for various tasks (e.g., pipeline validation, systematics studies). Each survey would need to appoint a researcher who has easy access to the working group requirements. The outcome from Phase 1 would be a comprehensive report that details the outputs from the simulation, analysis tools (e.g., halo finding approaches, ray trace code implementations), and how these are connected to the different survey tasks. 
\item \textbf{Phase 2: Tool development, validation, and cross-comparison; $\mathbf{\sim}$12 months}\\
During the second phase, all tools identified in Phase 1 have to be implemented and validated. Cross-code comparisons would be extremely useful. In addition, conversion schemes and readers for the different codes would be developed to enable sharing of the different data products in a straightforward way. \cite{2018ApJS..234...36M} demonstrate how this can be achieved across different synthetic catalogs -- the same approach, a reader that takes any input and converts it into a common exchange format, would be applied to enable sharing of data products between the surveys. Given that the tools need to be developed to run at scale and some of them to run on-the-fly within the simulation codes, the second phase will require considerable effort. We emphasize, however, that most cosmology codes already have at least a subset of the tools available. 
\item \textbf{Phase 3: Implementation; $\mathbf{\sim}$2021-TBD} \\
During the third phase, new simulations would be carried out that can  easily be shared between the surveys, given the preparations during the first two phases.  The computing resources needed for this phase are considerable  and would likely need to be obtained via competitive processes, such as INCITE in the US or PRACE in Europe. The effort required to run the simulations and enable the sharing of the associated data products strongly depends on the number of simulations to be carried out.
\end{itemize}

\subsubsection{Hydrodynamical Simulations}
The simulations and tasks mentioned below in Phases 1 and 2 fall under the Agencies' pre-existing research and development models for numerical projects. In addition, the Agencies' existing grant and award solicitations are sufficient to support the efforts highlighted. However, we recommend that the Agencies emphasize such proposals in grant programs including, but not limited to, NSF-AST, NASA-TCAN, NASA-ATP, and DOE and NSF Career awards. We also encourage the Agencies to fund multiple proposals in these solicitations to diversify the code development, subgrid modeling, and comparison efforts. The initial funding selection for such efforts is critical to begin as soon as possible to have new subgrid models tested and implemented. These hydrodynamic simulations are essential for the systematic mitigation and cross-correlation measurements for Rubin and Euclid, thus need to be completed by the time of first light for these surveys. A second round of funding will be necessary to further develop subgrid models for WFIRST and to update them with the new observations and tests provided by Rubin and Euclid.
\begin{itemize}
\item \textbf{Phase 1: Calibration of subgrid models; $\mathbf{\sim}$12 months}\\
As stressed in the report, a major challenge for hydrodynamic simulations in the cosmological context is the calibration of subgrid models. The work required in this area involves major R\&D efforts that are, at this point, not necessarily tailored to the specific surveys but still rather generic due to the large uncertainties in the modeling. Nevertheless, lessons learned should be shared between different groups and a concerted effort that enables easy sharing of results would be extremely beneficial. In addition, a comprehensive list of validation data sets, tests, and criteria relevant for the three surveys should be compiled. This list would be shared between the surveys and be used as a benchmark for the subgrid model implementations. In addition, it would have to be ensured that the major R\&D efforts are sufficiently funded to continue their efforts on developing and improving  the current subgrid models in different codes.

\item \textbf{Phase 2: Initial Model Implementation; $\mathbf{\sim}$36 months}\\
During the second phase, a range of simulations would be carried out that would be shared between the surveys. An important aspect here is that different approaches and subgrid model implementations would be automatically compared -- a major advantage of a cross-survey effort given the resources these simulations take. 
The simulations runs would be bracketing the remaining uncertainties in
the subgrid models. This conservative approach would capture the systematics associated with
baryons on the various cosmological estimators pertinent for the
surveys as discussed later in the report. The coordination between the surveys and the comparison effort 
would constitute in a multi-year program. An optimistic estimate would be 3 years of effort to actually make sufficient progress with this task. 
General support for this effort will enable different groups to scale up their codes to take full advantage of the planned exascale machines that are expected to arrive in $\sim$2021.

\item \textbf{Phase 3: Large Hydrodynamic Simulations; $\mathbf{\sim}$2022-TBD}\\
During the third phase, some groups might be in the position to carry out one or several large hydrodynamical simulations in cosmological volumes with sufficient mass resolution to undertake detailed studies of, e.g., intrinsic alignment effects and cluster physics. Such simulations would require major computational resources, which could become available in the U.S. in $\sim$2021. As part of a cross-survey activity, these simulations would be set up in a way that they could be easily shared across the surveys to address a range of questions with regard to baryonic effects on cosmological observables. The details of such a simulation campaign are currently too uncertain to outline here and will depend critically on the development and success of the U.S. exascale program (and in the more distant future the European exascale program). Due to these uncertainties, it is difficult at this point to estimate any resource requirements.

\end{itemize}


\section{Large Simulation Campaigns}
\label{sec:campaigns}

\subsection{Introduction}
Linking measurements of upcoming surveys to physical model parameters requires
very demanding forward simulations, which evolve the universe from early times to the present day. Extracting precision cosmological information from surveys depends upon extending existing modeling capabilities further into the small scale nonlinear regime as well as rigorous marginalization over currently unknown physics. In practical terms, it means that no single simulation can be sufficient for inferring new cosmological insights from observations, but that large simulation campaigns producing ensemble runs, while varying cosmological and other parameters, are needed. While no simulation in the ensemble will be at the level of the extreme-scale simulations discussed in the previous section, they
are still computationally costly and require significant allocations on modern supercomputers.
Data produced by those ensemble runs can be many petabytes in size, matching or even surpassing the data volume produced by the extreme-scale numerical simulations.
In addition, the analysis of the suites of cosmological simulations is complex if the aim is to directly compare or apply them to the analysis of the observational data. 

Two concrete examples of cosmological probes that will depend in the future crucially on accurate predictions in the nonlinear regime are weak lensing and cluster cosmology. 
To exploit the potential of the next generation of weak lensing surveys, producing
accurate predictions of the matter power spectrum is critical.  The signal-to-noise ratio of the cosmic shear signal is highest on angular scales of 5-10 arcminutes, which corresponds to physical scales of $\sim$1~Mpc. The observed two-point ellipticity correlation functions are highly sensitive to small scale structures projected along the line-of-sight, which means that restricting analyses to large scales is not a viable approach.

Currently, N-body simulations allow us to accurately capture the effects of non-linear structure formation on small scales and the requirement for a survey such as Euclid is to achieve a matter density power spectrum accuracy at the $\sim$1\% level.  This requirement goes beyond
the $\pm$5-10\% uncertainty of the popular Halofit code by \cite{2003MNRAS.341.1311S} with updates from \cite{2012ApJ...761..152T}. However, \cite{2009ApJ...705..156H} was able to recover a matter density power spectrum accuracy
of $\sim$1\% on scales out to k $\sim$ 1~Mpc$^{-1}$ for gravity-only simulations using Gaussian process modeling and sampling a five dimensional parameter space at only 36 points.
It is important to distinguish between gravity-only simulations, which are used to make the forecasts for cosmological surveys, and hydrodynamical simulations, which are able to describe modifications to the gravity-only matter power spectrum arising due to baryonic physics. Although baryons represent $\sim$1/6th of the total matter content in the universe, the distribution of baryons predominantly (at scales relevant here) traces the underlying dark matter density field and thus gravity-only simulations should capture most of the structure formation process. Nonetheless, differences in the spatial distribution of baryons with respect to the dark matter is expected to lead to changes that exceed the required accuracy in the matter density power spectrum of $\sim$1\% on small scales.

The primary summary statistic of cluster cosmology is the redshift-dependent mass function, i.e., number density of clusters as a function of mass and redshift.
Recent work by \cite{2018arXiv180405866M} emulates the dark matter halo mass function in a 7-dimensional parameter space ($\Omega_m$, $\Omega_b$, $h_0$, $n_s$, $\sigma_8$, $w$ and $N_{eff}$) sampling only 40 points in a 4$\sigma$ range around the current best guess ``concordance'' cosmology.  \cite{2018arXiv180405866M} find that their emulation of the dark matter halo mass function is sub-percent accurate and already sufficient to serve the needs of the first Rubin  data release.  Going forward, beyond year 1 of Rubin, this halo mass function emulator will need to be rebuilt with more accurate dark matter only simulations -- and likely more evaluated points in parameter space -- but it certainly appears that our ability to quantitatively describe the number density of halos as a function of mass and redshift is unlikely to be a bottleneck in future data analysis.

However, accurate cluster cosmology also critically depends on the knowledge of the
mass-observable relation and its scatter, which must be extracted from simulations for a wide range of cosmologies.
In an actual survey, clusters are binned by an observable that is correlated to cluster mass, for example redMaPPer richness, X-ray luminosity or temperature, or Sunyaev-Zeldovich signal. Presently, this cluster mass calibration error dominates the (theoretical) error budget, and is likely to be the main roadblock for cluster cosmology in future, possibly requiring the use of hydrodynamical simulations to ultimately resolve.

Beyond obtaining accurate predictions for a range of cosmological probes, 
measuring constraints on cosmological parameters relies on sampling schemes, such as Markov Chain Monte Carlo (MCMC) in order to explore likelihoods in parameter space.  Future surveys require tens of cosmological and nuisance parameters, and one needs to sample millions of different points in parameter space to reach convergence. Running a full ab initio cosmological simulation at each point in parameter space is thus not a practical solution, even if highly (physically) approximate methods would suffice. Investigations of advanced alternatives is a very active field of research, and includes algorithms for optimal sampling of parameter space and the interpolation of the target summary statistics given some sparse sample of evaluated points in the parameter space. This approach (colloquially called ``emulation'') was first introduced for the matter density power spectrum in \cite{2006ApJ...646L...1H}, and was more recently followed by the work on the halo mass function (\cite{2016ApJ...820..108H}, \cite{2018arXiv180405866M}), galaxy clustering and galaxy-galaxy lensing by \cite{2017arXiv170907099W}, and galaxy power spectrum and correlation function (\cite{2015ApJ...810...35K}, \cite{2018arXiv180405867Z}).

The examples listed above -- accurate predictions across cosmologies for a range of cosmological probes, investigation of different baryonic feedback models -- and also the need for covariance estimates, all showcase the need for generating ensembles of simulations. As for the extreme-scale simulations, results from such efforts can and should be easily shared between the different surveys. In particular, no survey specific modeling is required when building, e.g., emulators and therefore results are easily usable by a range of surveys.

\subsection{Key Challenges}

There are several challenges connected to generating large simulation ensembles. Some of these challenges are the similar to those for the extreme-scale simulations, but additional challenges arise due to the complexity of handling and organizing a large number of simulations. As for the extreme-scale simulations, securing computational resources, allocations as well as storage, to enable the runs themselves is difficult. However, the advantage is that each individual simulation is relatively small, so many more supercomputing facilities can be engaged to carry out such simulations. At the same time, if one wants to  take full advantage of a range of computing resources, there are major challenges for the simulator related to running and monitoring the simulations across multiple facilities.

The major challenges for carrying out large ensemble runs are:
\begin{itemize}
\item Securing computational resources (allocations, storage) to enable the runs themselves.
\item Developing analysis tools to efficiently extract a range of measurements from the simulations to enable the construction of emulators.
\item Building workflows that enable management for running and analyzing very large numbers of simulations (potentially across multiple facilities with varying architectures and requirements).
\end{itemize}

\subsection{TACS Findings for Large Simulation Campaigns}
Producing ensembles of simulations that span cosmological and nuisance parameters is essential to be able to fully exploit the information available from future cosmological surveys.
At this point, only a few such emulation projects have been carried out, mostly focusing on statistics that are easily extracted from N-body simulations, such as the matter density power spectrum and more recently galaxy-related statistics.
In future, those emulators closer to direct observable statistics will become crucial, including galaxy (photometric)-shear, galaxy-galaxy (photometric) correlation, shear-CMB cross-correlation, shear-CMB lensing cross-correlation, galaxy-CMB cross-correlation, and others.
The main difficulty in going from matter to galaxy statistics is the increased number of parameters, although nuisance parameters describing galaxy-halo relations need not be sampled with expensive ab initio simulations, but can be included in post-processing. The key to a large simulation campaign succesfully addressing the needs of multiple surveys is therefore separating parameters into computationally ``expensive'' and ``cheap''.  Cosmological parameters belong to the first group; changing any of them requires running a new simulation starting from linear-theory initial conditions.  Cheap parameters are, on the other hand, straightforward to vary directly on outputs, which can be done in post-processing.

The demand on the numerical codes is less severe here than with regard to the extreme-scale simulations; as ensembles consist of medium- to low-resolution simulations, code scalability is not as much of an issue, nor is I/O efficiency as each file is moderate in size.
However, we emphasize that future supercomputing architectures (beyond 2020) are anticipated to be more complex, thus current ``workhorse'' codes like Gadget-2 will not suffice unless properly modified.

\subsection{Conclusions}

We conclude the section by presenting common actions in support of large simulation campaigns useful for surveys considered here. 

\begin{itemize}
\item \textbf{Phase 1: Designing the common ensemble of simulations; $\mathbf{\sim}$12 months}\\
The goal of this phase is to obtain a comprehensive understanding of the needs of all science working groups in major surveys, and to design a minimal common grid of simulations.  This necessitates interactive collaboration involving scientists with strong expertise in different probes which rely on simulations for producing theoretical backdrop against which the observations are interpreted.  First challenge is to understand minimal simulation requirements needed to emulate different statistics at the desired level of precision. This commonly requires lot of domain knowledge, as the only way to access this information on parameter sensitivity without running full simulation ensemble itself is via approximate models.  In some cases, like the halo mass function, an analytic fitting function can be used as the approximate model.  But in other cases it relies on extrapolation of the accuracy of emulators build via coarser simulations.

This phase requires approximately one calendar year 
with researchers whose expertise spans the range of cosmological probes covered in this section.  This phase will also involve running some ``cheaper'' ensembles of simulations, but it would not be very computationally expensive, roughly millions of CPU hours.

\item \textbf{Phase 2: Ensemble production; $\mathbf{\sim}$24 months}\\
The goal of this phase is to produce the simulation ensemble, and it can start only after the successful completion of Phase 1.  The work involves proposing computational resources, managing the simulations on HPC platforms, assessing the emulation accuracy, adding new simulation points as mandated by the target accuracy.
This phase requires approximately two calendar years. 
Unlike Phase 1, most of the work will be very computational in nature. 
It is important that the scientists involved have some level of expertise in the field in order to tackle issues related to the final accuracy. The computing resources required for this phase are high and would likely have to be secured via competitive allocation process such as ALCC, INCITE or PRACE.

\item \textbf{Phase 3: Updates and support with designs involving non-cosmological parameters; $\mathbf{\sim}$24 months}\\
The goal of this phase is to provide support for the individual surveys that would use the common ensemble of simulations to add nuisance/post-processing parameters, like those needed to populate dark matter halos with simulated galaxies. We stress that the labor involved in this phase is not only data curation, but is iterative interaction with the relevant science working groups in the surveys.  As different statistics may need to have higher accuracy at certain points of the N-dimensional parameter grid, new point evaluations (i.e. full simulations) would be needed.  Once those are produced they could be propagated to other surveys and working groups as this would represent an overall increase in the modeling accuracy of the emulator.
This phase would continue for roughly 2 years.

\end{itemize}

\section{Generation of Synthetic Sky Maps}
\label{sec:maps}

There are a wide variety of methods for producing synthetic sky maps and there are many parallel efforts currently underway (using the same base simulation in many cases). We report on the approaches by the different surveys for generating synthetic sky maps, including the modeling of different galaxy types, generation of shear maps, validation approaches, etc. Common modeling and validation challenges have been identified and possible joint solutions and pipelines will be outlined.

Methods for generating synthetic skies for cosmological surveys can be loosely stratified according to modeling choices driven by the tradeoff between complexity and computational efficiency. We begin by briefly summarizing the broad categories into which contemporary methods fall, listed in descending order of the computational expense to generate a single synthetic sky:

\begin{enumerate}
\item {\bf Hydrodynamical simulations} of cosmological volumes \citep[for a recent review article, see][and references therein]{somerville_physical_2015} directly track the evolution of gravity-only particles such as dark matter, simultaneously with the physics of baryons, including fine-grained ``sub-grid'' prescriptions for processes such as radiative cooling, star-formation and associated feedback, black hole activity, etc. 
\item {\bf Semi-analytic models (SAMs)} \citep[for a recent review article, see][and references therein]{somerville_physical_2015}  are grafted into gravity-only N-body simulations. As a prerequisite to generating a synthetic sky, all SAMs require a significant post-processing phase of such N-body simulations, in which dark matter halos are identified at each output simulation timestep; halos across timesteps are subsequently linked together into a ``merger tree" that stores the evolutionary history of each identified halo. The SAM approach is to parameterize baryon-specific processes as functions of the halos and their evolution; on a halo-by-halo basis, SAMs seek to directly model how baryons would have been evolving had they been included in the N-body simulation. 
\item {\bf Empirical models} are also grafted into N-body simulations. All empirical models require the identification of dark matter halos, though the level of detail of the post-processing halo-identification phase varies considerably from method to method. These models are statistical in nature, as they are formulated in terms of stochastic mappings between ensembles of halos and ensembles of galaxies \citep[for a recent review article, see][and references therein]{wechsler_tinker18}.
\item {\bf Approximate N-body methods} employ various analytical techniques to circumvent the need for a full simulation. Some methods approximately solve for the evolution of the density field, and then identify halos in a post-processing phase; other methods approximately solve for the halo distribution more directly, without appeal to a halo-finder. All approaches require supplementation from simplified empirical models for the galaxy-halo connection to compute cosmological observables of galaxies from the approximated halo distribution. 
\end{enumerate}

Current and planned large-scale structure surveys most commonly employ empirical modeling and SAMs in the generation of synthetic skies supporting the survey. While hydrodynamical simulations are used extensively to study the impact of a variety of systematic effects that are relevant to large-scale structure cosmology, these simulations are seldom used to directly produce mock catalogs for collaboration-wide analysis.  Catalogs based on approximate N-body methods are typically used in applications where a single tracer population is distributed across a large cosmological volume; thus synthetic galaxies in mocks generated with these methods have essentially no attributes (beyond being brighter than a single color-magnitude threshold).  However, many scientific analyses involve making a range of cuts in multiple wavebands, which requires mock galaxy catalogs to have more complexity than has yet been achieved via approximate N-body methodology.

\subsection{Required Predicted Properties}

In this section we enumerate the properties typically required for simulated (imaging) surveys and assess the ability of each modeling approach to provide these properties. 

\subsubsection{Galaxy Flux, Color, and Stellar Continuum Spectral Energy Distributions (SEDs)}\label{sec:ss:broadband}

The broadband flux of a galaxy is one of the most important quantities required by any simulated sky program. Current and planned imaging surveys are composed of five or more filters, and the distribution of observed galaxies in this multidimensional space exhibits a rich spectrum of correlations across redshift and environment. Cosmological surveys have diverse needs for mock catalogs with accurate conditional one-point functions between most or all bands of the survey. Many cosmological analyses additionally require or benefit from mock catalogs with high-fidelity two-point functions, including correlations with color, brightness, and redshift. This is especially challenging because of the sensitivity of two-point clustering to environmental correlations, and because in all present-day methods these correlations are not parameterized directly, but emergent. 

Generating large-volume, synthetic galaxy catalogs that meet these specifications is a difficult challenge for all approaches to the problem. Traditional empirical models only produce mock galaxies with stellar mass or absolute restframe magnitude in a single band. By itself, such a model is insufficient to generate the required properties, and so empirical methods tend to be used primarily as  ``baseline'' or ``tuning'' mocks, on top of which additional modeling is carried out. The one- and two-point fidelity of mocks produced in this fashion is the highest of any available alternative, but with restricted applicability to the particular bands used in the tuning. Variations on this multi-step empirical approach are widely used to generate present-day catalogs; scientists within each survey typically develop one or more of such methods themselves using methods that commonly remain proprietary. 

SAMs use stellar population synthesis (SPS) models to produce a stellar continuum SED for each galaxy; from this, absolute magnitudes in multiple bands are found by integration of the SED under the appropriate filter. This modeling formulation naturally lends itself to a broad range of surveys, since in principle mock observations with any set of filters can be made on galaxies in a SAM-generated mock. In practice, it is computationally expensive to train SAMs, and difficult to ensure high-fidelity reproduction of the observed one- and two-point functions. The computational expense of SAMs is especially challenging because, in practice, the validation criteria of large imaging surveys are continually evolving, so that expensive one-time-only model calibrations quickly become obsolete. 

In hydrodynamical simulations, just as in SAMs, the star-formation and assembly history of each galaxy is used together with an SPS model to produce an SED, together with the flux observed through any desired broadband filter. Hydrodynamic simulations offer the highest level of complexity in the fluxes of synthetic galaxies, and considerable progress has been made in realistic forward-modeling of galaxy colors using hydro simulations. However, as described in Section~\ref{subsubsec:hydro_assessment} above, the computational expense of this approach makes it difficult to attain the necessary level of accuracy in the multi-${\rm Gpc}$ volumes required by present and planned surveys. 

The observable luminosities and colors will be affected by internal dust extinction in each galaxy. Modeling the effects of dust ranges in complexity and realism from simple dust screen models, through idealized geometry models in SAMs \citep{2016MNRAS.462.3854L}, to full ray-tracing calculations applied to some hydrodynamical models \citep[e.g.,][]{jonsson06}. Full ray-tracing is likely to be computationally too expensive for simulation of large surveys, and in any case, may not give accurate results if the input galaxies are poorly resolved. Screen and idealized geometry models are clearly oversimplified, but may at least give plausible scaling of dust extinction with galaxy properties (e.g. metallicity, surface density), and can be calibrated to match observational constraints.

\subsubsection{Spectra}

For spectroscopic surveys it may be necessary to construct complete spectral energy distributions for each galaxy or, at the least, to model key emission lines which will be used to select galaxy samples. As described in Section~\ref{sec:ss:broadband} both SAM and hydrodynamical simulations ubiquitously produce stellar continuum SEDs as a necessary step to the production of broadband luminosities, although the resolution achieved by the underlying SPS model may be insufficient to meet the resolution requirements of some surveys. Furthermore, as noted above, calculation of the full SED may be significantly more expensive than computing a small number of broadband luminosities in some cases.

Incorporation of emission lines into the spectrum has typically been achieved by modeling HII regions (e.g. using Cloudy \citep{2017RMxAA..53..385F}, which can also produce HII region continuua if those are required), with the physical conditions (e.g. metallicity, ionizing spectrum) taken from the underlying model (e.g. SAM or hydrodynamical simulation). Applications of this approach have been successful in reproducing luminosity functions and redshift distributions of H$\alpha$-emitting galaxies \citep{2008MNRAS.391.1589O,2010MNRAS.405.1006O,2014MNRAS.443..799O,2018MNRAS.474..177M}, although achieving plausible line ratios has proven to be more challenging \citep{2018MNRAS.474..177M}. All of the caveats regarding the effects of dust extinction on stellar continuum light also apply to emission lines, but are likely to be even more important as the emission lines arise preferentially from dense, dusty regions of galaxies.

Incorporation of the AGN component into spectra is much less developed. Both SAM and hydrodynamical models usually predict the masses and accretion rates of the central supermassive black hole in each model galaxy. These can be coupled with empirical or theoretical models of accretion disk spectra to compute the AGN contribution to the spectrum \citep{2011MNRAS.410...53F,2012MNRAS.419.2797F}.

\subsubsection{Morphology Indicators}

Morphology, by which we mean both classical morphological features (e.g. spiral vs. elliptical), but also size and shape, is crucial for construction of realistic simulated images, and for assessing the viability of weak lensing science. The greatest demands on morphology come from weak lensing working groups, which typically require mock galaxies to have both size and internal structure such as ellipticity and surface density profiles, including reasonably realistic correlations with broadband flux, morphological type, and redshift. 

In semi-analytic modeling, the typical approach is to model morphology as a two-component disk/spheroid, based upon the formation and merging history of each galaxy. Sizes are usually determined from the angular momentum of halos for disks \citep[][but see \protect\citealt{2018arXiv180407306J} who argue that sizes are uncorrelated with halo spin and determined instead by halo concentration]{1980MNRAS.193..189F,mo_formation_1998}, and energy conservation arguments for spheroids \citep[e.g.][]{cole_hierarchical_2000}. Within the class of SAMs there is a broad range of complexity in size modeling (e.g. inclusion of self-gravity of baryons, adiabatic contraction, energy dissipation during mergers). No extant model provides information about the ellipticity of the spheroid component. For disks, the normal vector of the disk plane is usually unspecified \citep{cole_hierarchical_2000}, or related to the angular momentum vector of the host halo \citep{stevens_building_2016}. 

Empirical models of galaxy morphology are at a less mature stage relative to models of SED-derived properties. There have been a comparatively small number of such models \citep[e.g.][]{RB09,SBM+09,DW17}; no published empirical model has attempted {\em joint} predictions for morphology together with SED-derived properties such as broadband color. 

Modeling the intrinsic orientation of galaxy morphology is worthy of special mention in the context of weak lensing science. Although intrinsic galaxy alignments are one of the leading systematics in lensing-based cosmological inference (see Section~\ref{sec:IA}), models of this effect in synthetic catalog generation are immature. This situation is partly due to limited availability of clean observational measurements of intrinsic galaxy alignments, but is primarily driven by the demands on modeling complexity created by the need for realistic covariance between orientation, morphology and SED-derived properties. 

The most detailed forward modeling of galaxy morphology in the published literature has been carried out using hydrodynamical simulations, which are coming to play a central role in studying intrinsic alignment systematics. However, in many respects the status of morphology modeling mirrors the situation reviewed in the previous section on SED-derived properties. While hydro simulations achieve greater complexity than models based on gravity-only simulations, their computational expense has thus far limited their direct use in synthetic catalog generation. 

\subsubsection{Clustering}

While information on the spatial distribution and clustering of galaxies is provided by the underlying dark matter distribution (provided by the base simulation), and is not a property of the galaxies themselves, it warrants mention here. Any measurement of clustering will involve some observational selection (e.g. on absolute magnitude), and so correlations between galaxy properties and their spatial distribution  must be correctly produced by models. Such correlations could easily be lost if galaxy models do not capture the details of assembly bias from the base simulation, or if the pre-processing of the base simulation does not capture sufficient detail (e.g. if it misses populations of subhalos, or fails to construct sufficiently accurate merger trees; \citealt{benson_convergence_2012,srisawat_sussing_2013,avila_sussing_2014,wang_sussing_2016}).

Furthermore, clustering predictions will depend to some degree on choices made in the treatment of sub-resolution effects by any given model. For example, \cite{2018MNRAS.474.5206K} explore the effects of the choice of how to model ``orphan'' galaxies (galaxies whose host subhalo is no longer detected in the base simulation, possibly for purely numerical/resolution reasons) affects predictions for two-point correlation functions, showing that these choices can have a significant effect on the amplitude of the two-point correlation function (and, therefore, on determinations of galaxy bias) due to the dependence of the frequency of orphan galaxy occurrence on the mass of the host halo. Validation of models in this respect must consider measures of clustering conditioned on a variety of observational selections.

\subsection{Key Challenges}

While techniques for survey simulations have advanced significantly over the past decade, there remain several challenges which must be overcome before any of these methods can meet the scientific requirements of forthcoming surveys. 

Primary among these challenges is that of calibration and validation. For SAMs (and hydrodynamical simulations), calibration is crucial to ensure that the models accurately match the target data. For empirical Monte Carlo methods, validation is key to demonstrate that the methodology is reliably robust.

In both cases, these requirements are strongly limited by the computational challenge. For SAMs, this challenge may be tractable \citep{henriques_monte_2009,lu_bayesian_2011,lu_bayesian_2012,bower_parameter_2010} using MCMC techniques, depending on the diversity of calibration datasets and the accuracy to which they must be matched, but will likely have computational expense of comparable order to that used to carry out the base simulations. Furthermore, while MCMC is efficient at searching the model parameter space, there is no guarantee that a viable model (one which matches the target data to within the required tolerance) exists within that parameter space. Such approaches also require careful consideration of the errors (both systematic and random) of each target dataset \citep{benson_building_2014}, including covariances---something which does not exist for the majority of datasets. The feasibility of calibration may also be limited by the validity of input physics modules. For example, it remains unclear whether extant SPS libraries produce colors to the required accuracy \citep{conroy_propagation_2010}. For hydrodynamical simulations, precision calibration is likely impossible on timescales of interest. 

Differences in validation criteria between projects may pose a challenge for synthetic sky models. All current models are imperfect, and are typically able to match only a subset of observational constraints simultaneously. If different projects have different validation requirements this may necessitate the construction of models tuned separately to each project---possibly invalidating any potential efficiency that could be obtained by utilizing the same model for multiple projects. Coordination on validation criteria between projects---with the goal of finding mutually-compatible criteria---should therefore be a priority. Additionally, observational constraints themselves are often inconsistent with other, similar constraints (e.g. two measurements of the galaxy stellar mass function which are formally different given their errors). Methods to allow for covariances between datasets, and systematic uncertainties in those data (as well as in the models themselves) have been explored \citep{2018arXiv180304470B}, but need further development to be applicable to the wide range of constraints and validation criteria that are expected.

The evolving nature of a survey's calibration requirements plays a critical and largely overlooked role in this challenge. A number of factors contribute to the evolution of these requirements as a survey progresses: additional scientists join the collaboration and bring new expertise that informs the criteria; contemporaneous surveys release new data or measurements; alternative analyses that complement the initially planned pipelines warrant new calibrations, and commonly require entirely new features of the model to be introduced. Our assessment is that this evolving nature is rather fundamental to the operating mode of all the large collaborations relevant to this report, and that this is unlikely to change for the indefinite future. The reason this aspect of the workflow is central to any discussion of the computational challenges involved in generating simulation-based synthetic skies is simple: the evolving nature of a survey's calibration requirements precludes the possibility of a one-time-only ``hero" calibration. This sharply contrasts with the challenges associated with running the N-body simulations. 

While the base simulations described in Section~\ref{sec:hero} are likely to dominate the computational cost of synthetic sky map production, the production and validation of galaxy populations will be a non-negligible and significant computational cost itself. The exact cost will depend on the strictness of the validation criteria for each specific survey, and on precisely which quantities are required (e.g. calculation of full SEDs is computationally much more demanding than producing just one or two broad band luminosities). As noted above, calibration and validation of models will almost certainly require performing multiple (likely $\gg 1$) runs of each model. Because of this, while calibration and validation can often be performed on a subset of the complete simulation volume, the total computational cost for calibration and validation is still expected to be of the same order as processing of the full simulation. Given current computational resources, calibration to the levels required is possible (though costly) for SAMs and empirical models, but impractical for hydro simulations.

Given these considerations, successful production of synthetic sky maps across surveys will require a significant investment of both computational and human resources. Due to the broad sweep of expertise across galaxy formation physics that is required to build a sufficiently complex and accurate model, and due to the evolving nature of the calibration requirements, the associated labor will need to be carried out in close collaboration with each of the surveys' analysis working groups.

\subsection{TACS Findings for Synthetic Sky Maps}

Synthetic sky map models are already able to meet the goal of being applied to the current generation of base gravity-only simulations, although in some cases this requires considerable computational resources. As the gravity-only simulations increase in volume and resolution, the sizes of the required synthetic sky maps increase, and the demands for modeling of additional quantities increase (particularly for multi-survey modeling), the computational demand of synthetic sky map production will increase significantly. While these demands will be met in part by the next generation supercomputers, significant investment of effort in code optimization, and development of statistical techniques to reduce computational demand will be crucial. 

Another significant challenge to be met is to produce synthetic sky maps which meet the requirements of science working groups---in terms of the diversity of galaxy properties which are modeled, the accuracy to which those properties match reality, and the extent to which key physical correlations between properties are captured by the model. Substantial efforts are needed to define validation criteria for models (particularly if they are to be used for multiple surveys where those criteria may be very different), to develop improved or extended modeling techniques where needed, and to develop more efficient methods to calibrate models. This labor cannot be effectively conducted by an individual or an isolated research group, but instead requires close collaboration with the survey(s) whose analysis working groups have needs for the synthetic catalogs. 

The funding structure of large cosmological surveys provides insufficient professional incentive to carry out this work. Currently, individual groups within a survey compete with each other to provide the synthetic mock that is singled out as the ``flagship'' or ``standard'' catalog of the collaboration; as generating mocks is a fairly specialized scientific activity, it is common for the graduate students and postdoctoral researchers involved to struggle to advance to the next career stage within the field. This competition-based funding model has thus far resulted in closed-source software packages with only modest applicability beyond the specific survey for which each package was tailored. 

Our assessment is that meeting the cross-survey goals outlined here requires a sustained effort to develop a scalable modeling platform with natural extensibility to multi-wavelength cosmological data. This platform would need to be developed in close contact with each survey's scientific working groups, and the code base would need to be open-source and adaptable to suit the needs of the specialized analyses within each survey. We consider it unlikely that any such framework will emerge in the absence of a new channel of stable, long-term funding dedicated to supporting the effort. 

\subsection{Conclusions}

We conclude this section by scoping the actions required by the challenge of generating synthetic galaxy catalogs that would be useful across surveys such as Rubin, WFIRST, and Euclid. 

\begin{itemize}
\item \textbf{Phase 1: Comprehensive Assessment; $\mathbf{\sim}$12 months}\\ 
Conduct a comprehensive assessment of the needs of all major surveys for whom the mock data is intended. This necessitates working closely with the relevant analysis working groups of each survey to build and achieve consensus on quantitative validation criteria that will be used to evaluate the mock. Special care must be taken to ensure that the validation data are self-consistent; each criterion should be associated with one or more specific science aims of the surveys. 

This phase could be accomplished in roughly one calendar year by 
scientists whose expertise spans the range of topics covered in this section. 

\item \textbf{Phase 2: Model Development; $\mathbf{\sim}$18-24 months}\\ 
Develop models and scalable software tools to generate galaxy catalogs with properties that are currently not available in mocks built for a particular survey. The end result of this phase is the formulation of a comprehensive model with sufficient complexity to meet each of the surveys' needs, and a scalable implementation that can efficiently leverage the architectures of leadership-class computing facilities. 

The labor for Phase 2 could in principle commence $\sim$3-6 months after the beginning of Phase 1 and we expect that Phase 2 could be completed 
 over $\sim$18-24 months. 
It is critical that the scientists involved have specialized expertise spanning the required range of fields, including semi-analytic and empirical modeling of the galaxy-halo connection, gravitational lensing (including generation of simulated lensing maps), software pipeline engineering, and data-intensive parallel computing.  

\item \textbf{Phase 3: Model Calibration; $\mathbf{\sim}$12-18 months}\\ 
Having built the form of the model and established quantitative optimization criteria, Phase 3 will result in the delivery of a synthetic catalog that simultaneously meets the needs of all surveys participating in Phase 1. We stress that the labor involved in Phase 3 is not merely managing a large computation, but will in fact be iterative with both of the previous phases: as the calibration effort proceeds, the scientists will discover new features required of the model, and the validation criteria will undoubtedly evolve as new data become available in the time spanned by this effort. 

Phase 3 necessarily follows Phase 2, and could be conducted 
over $\sim$12-18 months.

\end{itemize}


\section{Investigation of Astrophysical and Theoretical Systematic Effects}
\label{sec:syst}

This section reports on systematic effects that can be investigated by the surveys via the use of large-scale simulations. It is important to be clear about the scope of this section: We are not exploring the use simulations for systematics modeling in a general sense, in particular we are not looking at observational systematics such as photo-z and shear calibration, extinction, sky brightness. We stress that these effects also require the use of simulations, however they are more survey specific than synergistic and hence beyond the scope of this report. We do explore the use of simulations for modeling systematic effects that are not survey specific and where a joint simulation campaign would benefit each of WFIRST, Euclid, Rubin.

The systematics considered here include intrinsic alignments, baryonic effects, galaxy bias, non-linear evolution of structure formation, and projection effects. 
Most individual systematics tend not to cross-correlate, which is a strong incentive to investigate cross-correlations. This is particularly important given the fact that none of the astrophysical systematics are first-principle calculations (with the exception of perturbative galaxy bias expansions) but rather phenomenological descriptions that are based on observations and analytical approximations that are implemented through subgrid physics models or via semi-analytic models, the latter of which are added to the gravity only simulations in post-processing (also see Sect. \ref{sec:maps}). In this context it is vital to ensure an information exchange between ongoing observational campaigns (i.e. the Dark Energy Survey, Kilo Degree Survey, Hyper Suprime Cam Survey, Baryon Oscillation Spectroscopic Survey, and the Dark Energy Spectroscopic Instrument) that enhance our understanding of astrophysical models and the simulation campaigns of WFIRST, Rubin, and Euclid that need to implement the improved understanding derived from those surveys into increasingly refined simulations. This is an iterative process that requires close interaction of observers, theorists, and simulators, and it requires an equally close interaction of the research  frontier and large infrastructure efforts.   

\subsection{Accounting for Baryonic Effects} 

As optical and near-IR imaging surveys push the measurements of galaxy clustering and weak lensing into the non-linear regime, it is important to understand effects at smaller scales. In particular for weak lensing, the signal is mostly concentrated in smaller scales and thus accounting for baryonic effects on the matter power spectrum becomes critically important to provide an unbiased cosmological parameter inference \citep[e.g.][]{SHS13,ZSD+13,EKD+15}. This is also true for cluster science, where the need to characterize galaxy clusters with baryonic physics is becoming critical if one wants to use clusters to provide unbiased cosmological constraints (see, e.g., \cite{2016MNRAS.456.2361B} for a study of the impacts of baryons on the halo mass function).  There are currently multiple efforts to understand and simulate detailed baryonic physics within sizable ($\approx 100 h^{-1}$ Mpc on the side) cosmological volumes. A list of hydrodynamic simulations and their properties is given in Section~\ref{sec:hydro}, Table~\ref{tab:hydrolist}. A major focus of these studies in the cosmological context is trying to understand at which length scales (in Fourier space and real space analyses) baryonic physics become so important that predictions from gravity-only simulations cannot be used anymore for cosmological analyses. For a very recent comparison of different hydrodynamical simulations, including the EAGLE, Illustris, and IllustrisTNG100 and TNG300, see \cite{2018MNRAS.475..676S}, for their impact on weak lensing with future surveys a recent study can be found in \cite{2018arXiv180901146H}. 

At this point, more studies are needed to enable robust predictions for the matter density power spectrum on small scales and the effects of baryons on cluster mass measurements. Initiating a joint program across the surveys to tackle this question would enable detailed comparisons and studies of the (very different) subgrid physics models that are employed in these simulation efforts and how they affect the cosmological observables.

\begin{figure}
\begin{center}
\includegraphics[width=3.0in]{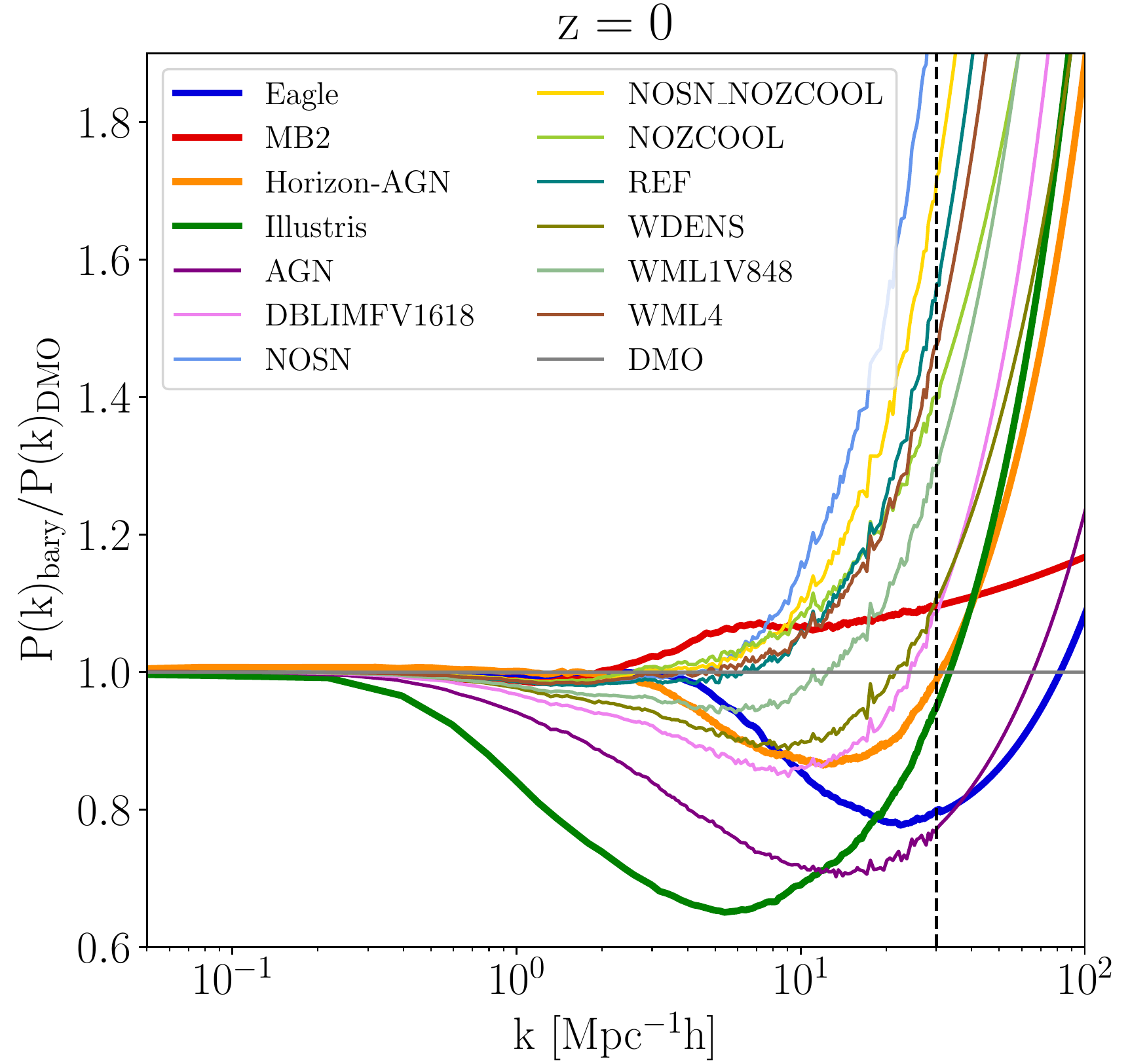}
\caption{The power spectrum ratio of different hydrodynamical simulations with respect to their counterpart dark matter only (DMO) simulations at $z=0$. The thick lines represent the cases for the EAGLE, MassiveBlack-II (MB2), Illustris, and Horizon-AGN simulations, while the thin lines indicate the 9 different baryonic scenarios in the OWLS simulation suite. The dashed vertical line divides the power spectrum ratios into regions where data points come from direct measurements (k $\le 30$ h/Mpc) or from extrapolation with a quadratic spline fit (k $\ge$ 30 h/Mpc). Figure taken from \citep{2018arXiv180901146H}.}
\label{fig:Pk_ratio}
\end{center}
\end{figure}

\subsection{Intrinsic Alignments}
\label{sec:IA}

Cosmic shear is typically measured through two-point correlations of observed galaxy ellipticities. In the weak lensing regime, the observed ellipticity of a galaxy is the sum of its intrinsic ellipticity, $\epsilon^{\rm I}$, and gravitational shear, $\gamma$: $\epsilon^{\rm{obs}} \approx \epsilon^{\rm I} +\gamma$. If the intrinsic shapes of galaxies are not random, but spatially correlated, these intrinsic alignment correlations can contaminate the gravitational shear signal and lead to biased measurements if not properly removed or modeled. Since early work establishing the potential effects \citep{hrh00, ckb01, cnp01}, intrinsic alignments (IA) have been examined through observations \citep[e.g.,][]{hmi07,jma11,bms12,smm14}, analytic modeling, and simulations  \citep[e.g.,][]{sfc12, tsm14,tmd14} - see \cite{tri14} and \cite{Joachimi15}, and references therein for recent reviews. A fully predictive model of IA would include the complex processes involved in the formation and evolution of galaxies and their dark matter halos, as well as how these processes couple to the large-scale environment. In the absence of such knowledge, analytic modeling of IA on large scales relates observed galaxy shapes to the gravitational tidal field and typically considers either tidal (linear) alignments, or tidal torquing models.

\begin{figure}[h]
\begin{center}
\includegraphics[width=3.0in]{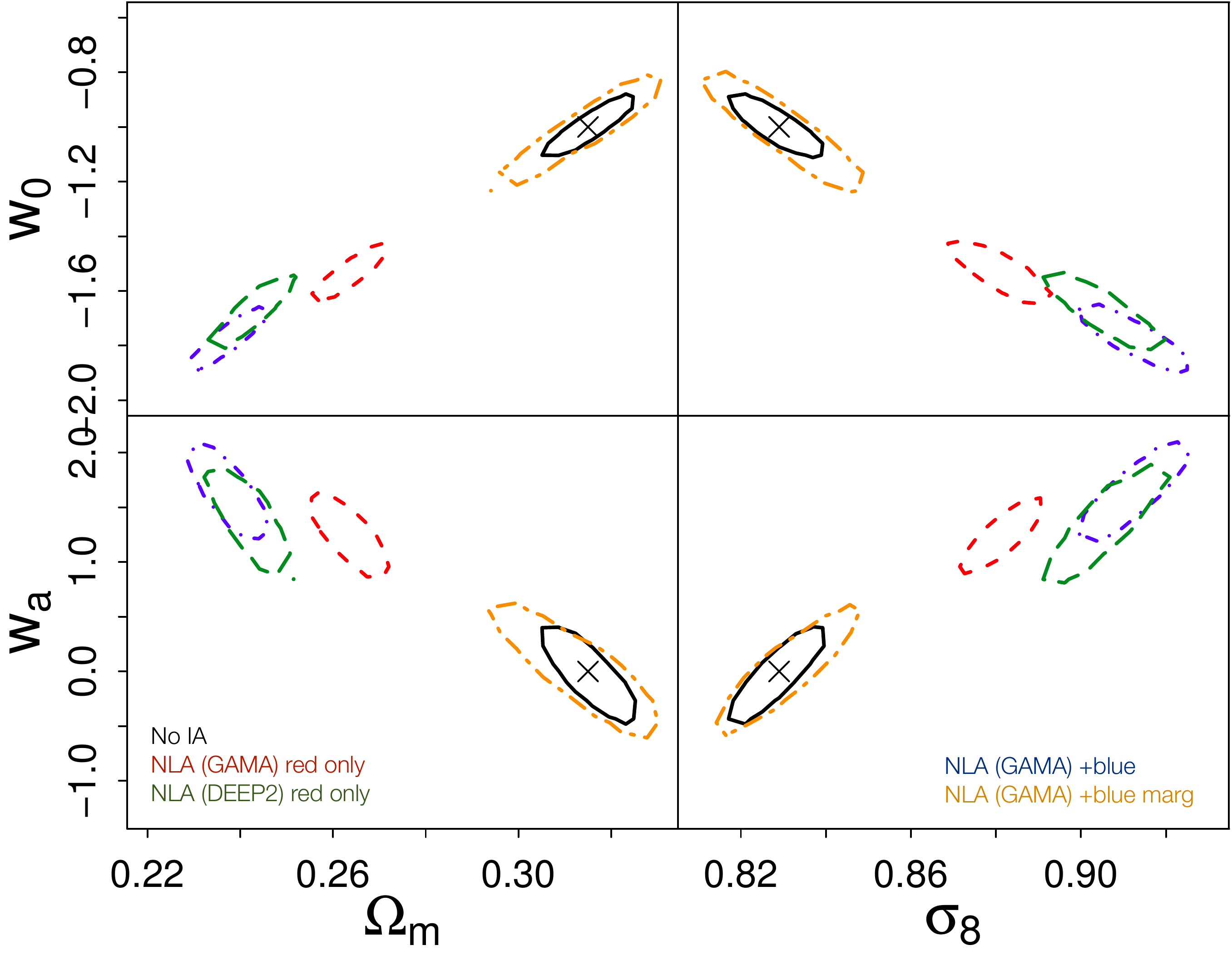}
\caption{The impact of IA on WL constraints (68 per cent confidence region) from Rubin assuming the nonlinear alignment (NLA) scenario. We consider different luminosity functions, i.e. GAMA (red/dashed) and DEEP2 (green/long-dashed) and for the GAMA luminosity function we also consider the case for which blue galaxies have a mild NLA IA contribution (blue/dot-dashed). The Rubin statistical errors are shown in black/solid. Orange/dot-long-dashed contours show results when using the most extreme of these cases, i.e. the data vector corresponding to the blue contours, as input and including a standard IA mitigation scheme in the analysis. The marginalized likelihood is obtained by integrating over a 11-dimensional nuisance parameter space (see text for details) Figure taken from \cite{2016MNRAS.456..207K}.}
\label{fig:IA_impact}
\end{center}
\end{figure}

The shapes of elliptical, pressure supported galaxies are often assumed to align with the surrounding dark matter halos, which are themselves aligned with the stretching axis of the large-scale tidal field \citep{ckb01,his04}. This tidal alignment model leads to shape alignments that scale linearly with fluctuations in the tidal field, and it is thus sometimes referred to as ``linear alignment,'' although nonlinear contributions may still be included \citep{brk07, bms11, bvs15}. For spiral galaxies, where angular momentum is thought to be the primary factor in determining galaxy orientation, IA modeling is typically based on tidal torquing theory, leading to a quadratic dependence on tidal field fluctuations \citep{ckb01, lep08}. However, on sufficiently large scales, a contribution that is linear in the tidal field may dominate. Due to this qualitative difference in assumed alignment mechanisms, source galaxies are often split by color into ``red'' and ``blue'' samples, as a proxy for elliptical and spiral types. Indeed, blue samples consistently exhibit weaker IA on large scales, supporting the theory that tidal alignment effects are less prominent in spirals \citep{flw09, hmi07, mbb11}. On smaller scales, IA modeling must include a one-halo component to describe how central and satellite galaxies align with each other and with respect to the distribution of dark matter \citep{scb10}.  \cite{2016MNRAS.456..207K} have conducted an exhaustive analysis of the impact of IA on Rubin weak lensing analyses (see Fig. \ref{fig:IA_impact} for some representative results), varying luminosity functions, IA models, mitigation schemes, and contamination fractions of blue and red galaxies. Numerical simulations, especially those including hydrodynamical physics, have recently become powerful tools for constructing these models \citep{sfc12, jsh13, tsm14, tmd14, 2017MNRAS.472.1163C}. It will be critical for the future to refine these simulations with the latest observations and to forecast the impact on Rubin, WFIRST, Euclid analyses, and to further refine this iterative approach to improve IA modeling. In this context it is of particular interest to study the correlations between IA uncertainties and galaxy-halo and baryonic modeling uncertainties and to develop a joint description of these intertwined astrophysical phenomena.

\begin{figure}[h]
\begin{center}
\includegraphics[width=3.0in]{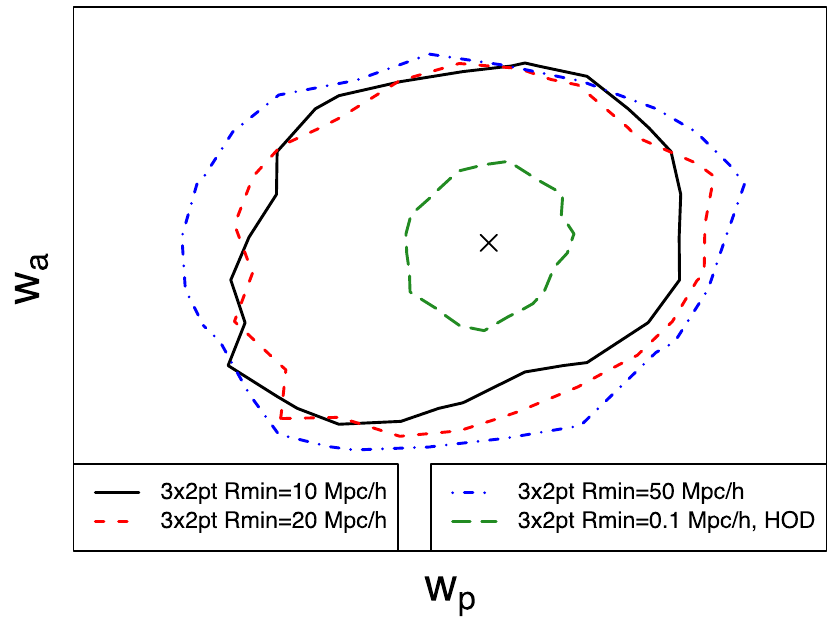}
\caption{This figure shows the gain in information on dark energy parameters $w_a$ and $w_p$ (where the latter corresponds to the commonly known $w_0$ parameter but computed at a pivot redshift $p$) as a function of varying the minimum scales included in galaxy clustering and galaxy-galaxy lensing measurements. We show results for an Rubin joint clustering and weak lensing analysis (so-called 3x2pt), which assumes $R_{min} = 10 \, \rm{Mpc/h}$ (black/solid), and corresponding constraints when using $R_{min} = 20 \, \rm{Mpc/h}$ (red/dashed), $R_{min} = 50\, \rm{Mpc/h}$ (blue/dot-dashed), $R_{min} = 0.1 \, \rm{Mpc/h}$ (green/long-dashed) instead. For the latter we switch from linear galaxy bias modeling to a 6 parameter Halo Occupation Density (HOD) implementation. Figure taken from \cite{kre17}.}
\label{fig:smallscales}
\end{center}
\end{figure}

\subsection{Going into the Nonlinear Regime}
\label{sec:galaxybias}
The nonlinear regime of structure formation holds a wealth of cosmological information. For Rubin this has been demonstrated in \cite{kre17} Fig. 4 (reproduced in Fig. \ref{fig:smallscales}). The figure shows the information content as a function of the minimum scales included in an analysis. For the black, red, and blue contours a standard linear galaxy bias model is assumed, whereas for the green contours, which include information from scales down to 0.1 Mpc/h, the analysis assumes a 6 parameter Halo Occupation Density model \citep[see][for details]{kre17}. This analysis is performed in a 50+ dimensional parameter space and we note that such a significant gain in information given, the high-dimensionality of the parameter space, is extremely rare.    
Rubin, WFIRST, Euclid have great potential to exploit this information if the scientists provide accurate predictions well into the nonlinear regime. This task is difficult -- not only due to baryonic physics that alter prediction on small scales but even to generate high-accuracy, gravity-only results across cosmologies is a difficult task. To this end, the nonlinear evolution of dark matter on large scales can be treated in different ways. One is using perturbation theory, which has been the default method when interpreting galaxy clustering in redshift surveys. It allows a somewhat more controlled understanding on semi-nonlinear scales. Another method employs phenomenological fits to N-body simulations based on the halo-model, like Halofit (for the most recent incarnation see \citealt{2012ApJ...761..152T}), or emulators of the actual N-body power spectrum measurements \citep{2017ApJ...847...50L}. A third approach is full forward modeling, where simulations are rapidly produced (using fast approximate codes) and comparing the outputs with observational datasets directly see, e.g., \citep{2017JCAP...12..009S}. A fourth approach includes using machine learning directly to predict cosmological parameters from the large scale structure to very small scales \citep{2017arXiv171102033R}. All of these methods need to be refined to reach the accuracy required for upcoming surveys. A joint effort to investigate the validity of these approaches, the most efficient implementation, and spatial reach at a given accuracy would be extremely valuable across the three surveys targeted in this report.

\subsection{TACS Findings for Systematic Effects}
Astrophysical systematics are common across all surveys and developing the required systematics mitigation strategies to optimize the science return of Rubin, WFIRST, and Euclid, requires an integrated effort that includes simulations, observations, and analytical descriptions.
Observations from precursor surveys such as DES, KiDS and HSC (in combination with CMB and spectroscopic surveys like BOSS and later DESI) provide information on e.g., galaxy bias, velocity bias, baryonic scenarios, and intrinsic galaxy alignments. Some of these systematics can be modeled through analytical expressions, which are then incorporated into numerical simulations in two ways: 1) via a post-processing step of N-body simulations or 2) through fine-tuning the sub-grid physics in hydrodynamical simulations. The increased precision of these simulations will in return enable an improved interpretation of Rubin, WFIRST, and Euclid data. This will be an iterative process, necessary nevertheless to avoid being dominated by astrophysical systematics in future surveys. It is important in this iterative process not to double-count information, i.e. to develop a thorough procedure such that the data used to improve the simulations is not also the data then analyzed with said improved simulations. Ensuring that the information used in the systematics simulations remains independent of the Rubin, WFIRST, Euclid data is critical.

\subsection{Conclusions}
TACS finds that teams should be selected through competitively selected grants that include experts from observations, simulations, and analytical modeling. These teams should be cross-institutional and cross-survey; they should include experts from precursor surveys (e.g., BOSS, DES, eBOSS, HSC, KiDS, SDSS, VIPERS, etc) and from external data sets (e.g., CMB, X-ray, SZ), experts on numerical simulations and analytical modeling. In this context it is important to note that astrophysical systematics are correlated with another and with cosmological observables and that developing strategies for each of the systematics independently will have very limited success.

\begin{itemize}
\item \textbf{Phase 1: Joint-probe, joint-survey assessment and forecasting; $\mathbf{\sim}$12 months}\\
Systematics modeling and mitigation experts together with numerical simulation experts from Rubin, WFIRST, Euclid, should share information on modeling strategies, existing/planned simulations, anticipated access to external data sets. The impact of the systematics (all of them together) for a joint-probe and joint survey analysis needs to be assessed properly (cosmological forecasts).  

\item \textbf{Phase 2: Calibration, Validation, and Verification; $\mathbf{\sim}$12 months}\\
Continuation of the forecasting effort but informed by early results from the simulations. The simulations should be calibrated with target observations and should be made as realistic as possible. At this point in time it should be clear what the relevant parameters in the simulations (e.g., subgrid physics) are that have the largest impact in changing the observables. This phase should also include developing a strategy to finetune the simulations via observables that are weakly dependent or if possible fully independent of cosmology.

\item \textbf{Phase 3: Systematics mitigation implementation; $\mathbf{\sim}$12 months}\\
Implementation of the systematics mitigation strategies into cosmological modeling frameworks. Simulations at this point should span a large range in realistic cosmological and systematics models and they should not violate any cosmology independent observables within reasonable error bars. Test of mitigation strategies using precursor data and using independent simulations. 
\end{itemize}


\section{Advanced Statistical Methods}
\label{sec:stats}

The analysis of cosmological data and simulations relies on using the most sophisticated statistical methods available today. The input to many of these methods are large numbers of dark matter-only simulations, as discussed in Section \ref{sec:campaigns}. Due to the high cost of these simulations, it is very important to study statistical methods that help to reduce the number of required simulations. Examples include the development of emulators (predictions tools) from a limited set of high-quality simulations spanning a range of cosmological parameters or new modeling techniques for covariance estimates to reduce the number of realizations needed. We have identified common statistical challenges that rely on expensive simulations and discuss possible alternative methods that should be evaluated further. 

\subsection{Next-generation emulators}

The creation of each virtual universe --- for a given set of cosmological and marginalization parameters, as well as the particular random realization of the initial density fluctuations --- requires an extremely computationally
expensive simulation on High Performance Computing resources. In order to make cosmological inverse problems practically
solvable, constructing a computationally cheap surrogate model or an emulator is imperative.
Current approaches to emulators require the use of a summary-statistic which is to be emulated, and are using simulations of the same fidelity for each ``design'' point in N-dimensional parameter space.

To meet future survey requirements, we expect next the generation of emulators to exhibit progress in the following ways: (1)  to have an iterative instead of a fixed design; (2) to be multi-fidelity capable, meaning to combine simulations done at different fidelities; and (3) use multi-level emulation via separating design into ``expensive'' (e.g. cosmology parameters) and ``cheap'' parameters, like those appearing in post-processing runs, responsible for predicting different luminosities or galaxy types from the density field.

\subsection{Covariances and Likelihood functions}
Methods to obtain covariances can be broadly structured into 3 different categories: 1) analytic covariances, 2) covariances estimated from numerical simulations, and 3) covariances estimated from the data directly. These methods have different advantages and disadvantages; precursor surveys of Rubin, WFIRST, Euclid have mostly been focusing on analytic covariances \citep{kez17,2018arXiv181002322A, 2017MNRAS.465.1454H, 2018arXiv180909148H} and only rarely on simulation based covariances. Covariances directly estimated from the data have not been used recently due to known biases when estimating the variance of a large survey size from smaller subsets (e.g., \citealp{2016MNRAS.456.2662F}). All analyses that used analytic covariances have had some validation scheme that involved numerical simulations.

Analytic covariance matrices have 3 main advantages: 1) computational feasibility for large data vectors, especially in multi-probe analyses \citep[see e.g.][for a 7+million entry joint covariance of weak lensing, galaxy-galaxy lensing, galaxy clustering, cluster number counts, and cluster weak lensing]{kre17}, 2) simple inversion procedures, and 3) flexibility in terms of the scales, redshifts, galaxy samples that are considered. Whereas the second-order and Supersample variance terms can be computed sufficiently precisely (and much faster) using analytic covariances, the question remains whether higher-order moments of the density field are sufficiently precisely captured using analytic descriptions, primarily via the halo-model. In the case of weak lensing \cite{bks18} have recently shown that the Gaussian and Supersample covariance terms are dominant such that the higher-order (connected tri-spectrum) terms can be neglected, without biasing Rubin and Euclid likelihood constraints significantly. This result needs to be explored in the context of clustering, galaxy clusters and other probes, but for weak lensing it has become clear that analytic covariances are a viable solution.

When moving to simulation or data based covariance matrices, the scientific topic of choosing the best estimator is important. Extensive research has been conducted on covariances obtained via the standard sample variance estimator and on the impact of imperfect estimated covariances on the cosmological parameter constraints. At the heart of the problem is the simple fact that the Gaussian likelihood, which is commonly assumed in cosmological analyses, requires an inverse covariance. Unfortunately, the inverse of an estimated covariance is not the estimated inverse covariance and even minute residual noise in the covariance estimator can severely bias the inverse.
\cite{hss07} described a way to correct for this when assuming that the covariance estimate follows a Wishart distribution \citep[see also][]{Kaufman,and03}. The noise properties of this corrected precision matrix estimator and its impact on the constraints derived on cosmological parameters was e.g. investigated by \cite{tjk13}, \cite{dos13}, and \cite{taj14}, where the authors pointed out the enormous number of realizations required (of order $10^6$ or even $10^8$) to achieve an inverse covariance with an acceptable precision.

Recently new Hybrid estimators (combining analytic and simulations and data) have emerged \citep{2018MNRAS.473.4150F} and linear and nonlinear shrinkage estimators are being explored \citep{2017MNRAS.466L..83J,pos08}
which have substantially reduced these estimates and further reductions are possible via data compression. 

The functional form of the likelihood being a multivariate Gaussian has been questioned in the literature, mostly in the context of weak lensing \citep{2009A&A...504..689H,2013A&A...556A..70W, 2018MNRAS.477.4879S}, but the same argument holds for galaxy clustering, galaxy-galaxy lensing and other large-scale structure probes. The core argument is that summary statistics derived from a non-Gaussian field have no first principle reasons to follow a multivariate Gaussian likelihood. In the context of the CMB, a corresponding approximation for temperature and polarization second order statistics has been shown to not bias results, however the CMB field is substantially closer to a Gaussian compared to the late Universe large-scale structure observables. 

Alternative approaches such as estimating the likelihood from simulations directly, or utilizing likelihood free analysis techniques such as Approximate Bayesian Computation are still in their early phase of exploration and require targeted research funding to mature fully as alternatives. The necessity of abandoning the multivariate Gaussian likelihood function as an assumption needs to be established first. Currently the literature does not conclusively state whether this approximation fails at the level of precision for Rubin, WFIRST, and Euclid. For this exploration we recommend a staged process of analytic exploration, inexpensive simulations, e.g. FLASK \citep{2016MNRAS.459.3693X}, and subsequently $\sim 10^3$ high precision simulations (the necessary number will be more precisely determined during the first two steps).

\subsection{Developing Discrepancy Metrics:}
Developing meaningful discrepancy metrics is a core element of interpreting cosmological data. The most prominent questions are: Is model A preferred over model B (LCDM vs wCDM in the most simple case)? Is dataset A in tension with dataset B (Euclid vs WFIRST vs Rubin)? Before combining datasets, scientists must assess whether the data to be combined are in tension with one another in the context of a given cosmological and systematics model. 

Discrepancy metrics are also important for a joint simulation effort of Rubin, WFIRST, and Euclid namely in determining whether the simulations are sufficiently precise given the constraining power of the surveys individually and then jointly. This is not a trivial task since, in principle, such an assessment requires an even more precise simulation of the survey(s) in the first place. Even in the presence of such a fiducial high-precision simulation (e.g., see the Euclid Flagship simulation), the questions arise: what precision do the emulator simulations need, what precision do the covariance/likelihood simulations need, and what precision do the systematics simulations need?

The most common ways to quantify discrepancies are through either biases in the w0-wa parameter space that arise when piping an imperfect simulation through a survey simulation pipeline or through the increase in the error bars, usually also in the w0-wa parameter space, when accounting for an imperfect simulation by adding nuisance parameters to the survey simulation or by adding the uncertainty quadratically to the covariance. But even here the analysis choices can decide on the outcome of the discrepancy quantification and can determine whether a simulation is deemed sufficiently precise or not. Common analysis choices include: 1) how the covariances are computed, 2) what probes are included in the analysis, 3) what scales, redshifts, and galaxy samples are selected, 4) what priors are assumed from external data, 5) what systematics are included in the survey simulation, how are they parameterized and what are their priors, and most importantly 6) what physics is included in the parameter space (e.g., neutrinos, curvature, dark matter models).

In order to asses whether a simulation (campaign) is sufficiently accurate for the individual surveys Rubin, WFIRST, and Euclid and additionally for their joint analysis, it is critical to unify the analysis choices for the survey simulations across the Rubin, Euclid, and WFIRST communities. This will allow the surveys to have a meaningful framework to assess whether a given simulation (campaign) is suitable for their needs. It is furthermore important to not simply quantify the precision of the simulations for time-dependent dark energy, aka the w0-wa plane, but also for more general dark energy models and modified gravity scenarios (e.g., alpha parameterization).

Most importantly, the assessment of whether a simulation (campaign) is sufficiently precise should happen early, i.e. during the planning phase, of said effort. 

\subsection{TACS Findings for Advanced Statistical Methods}
Statistical methods are a critical element of a coordinated simulation effort across surveys. Corresponding research is indispensable in order to efficiently use the existing computational resources (examples are emulators and covariance estimators) and in order to ensure that simulations generated within one of the surveys are meaningful for another survey (discrepancy metrics). 

\subsection{Conclusions}
The pilot studies suggested below are best implemented through competitive research grants from the DOE, NSF, and NASA, or small ``Tiger Teams" that combine expertise in statistical methods and numerical simulations across the surveys. 
First and foremost the surveys should share expertise and code on the topics below and develop a coordinated testing scheme of the code implementations. The simulation resources required to implement some of the solutions on emulators, covariances/likelihoods, and discrepancy metrics should be shared and the solutions should be tested on these shared resources in Phases 2 and 3. For example, covariance estimators using simulations that are developed by each survey should be tested against one another in simulated likelihood analyses.   	 
\begin{itemize}
\item {\bf Emulators:} Emulation of the computationally expensive aspects of a survey analysis is an indispensable concept in survey cosmology. Research on improved emulators, especially in the context of new Machine Learning concepts, that interface statistical expertise with expertise of numerical simulators should be a priority in competitively selected grants. 
\item {\bf Covariances/Likelihoods:} 
\begin{enumerate}
\item The current state-of-the-art for covariance generation, given the latest results in analytical computation, hybrid estimators, and non-linear shrinkage estimators requires of order $10^3$ simulated survey realizations (previously $10^6$ or even $10^8$).  
\item Even $10^3$ simulation realizations is a pessimistic scenario. 
With continued investment, the required number of simulations could plausibly decrease even further (possibly to of order $10^1$) through a combination of data compression ideas and through combinations of the aforementioned estimators. 
\item The tolerable error in the precision matrix is also dependent on the systematics budget and the overall dimensionality of the likelihood analysis. In high-dimensional parameter spaces, errors in the precision matrix translate into sub-dominant uncertainties.
\item Covariance matrices are only required when assuming that the likelihood of the considered summary statistic is a multivariate Gaussian. This assumption breaks down, strictly speaking, in the context of estimated covariances (from simulations and/or the data directly), where it has been shown that the likelihood follows a t-distribution. Even in the case of an analytic covariance matrix, the commonly considered summary statistics (two-point statistics) are not strictly speaking distributed as multivariate Gaussians. Initial results differ on the importance of this effect, i.e. the non-Gaussianity of the likelihood at least when considering two-point statistics.   
\item Alternative inference techniques, e.g. Approximate Bayesian Computation or Bayesian Hierarchical Modeling do not require assumptions on modeling a likelihood function and/or covariances. The community is exploring these avenues increasingly and although major obstacles remain, it should be on the survey community's radar.
\item {\bf Findings:} 
\begin{itemize}
\item The impact of uncertainties in precision matrices should be examined by a joint task force of experts across the surveys. This includes implementing and testing the new estimators in a realistic survey specific context. Data compression should be included in this effort. The goal of this effort should be to determine the required simulation effort for covariances across all surveys. 
\item Non-Gaussian functional forms of the likelihood and alternative inference techniques are largely unchartered territory in terms of simulation needs for Rubin, Euclid, and WFIRST. Active research on these topics through competitively selected grants should be prioritized.
\end{itemize}
\end{enumerate}

\item {\bf Discrepancy metrics:} 
Discrepancy metrics are critical to assess whether the quality of simulations is acceptable in the context of individual Rubin, Euclid, and WFIRST analyses and their joint effort. TACS finds that a cross-survey collaboration effort should be created to develop these metrics in the context of realistic analysis choices for the individual surveys. 
\end{itemize}

\section{Common Infrastructure to Share Simulation Products}\label{sec:infra}

\subsection{Introduction}
Rubin, WFIRST, and Euclid are all looking at the same sky in a similar time-frame and they all have similar requirements for cosmological simulations. At the simplest level, it is a poor use of resources for the three surveys to produce largely redundant simulation suites individually. In addition,  there are only a limited number of people in the world with the expertise to produce extreme-scale cosmological simulations and synthetic sky catalogs and also only a limited number of supercomputing facilities with the resources available to produce extreme-scale simulations or large suites of cosmological simulations. Given these limitations it is challenging for the surveys to realize their cosmological simulation needs individually. In practice, it is the same simulators being approached by the different surveys with slightly varying requests for cosmological simulations and their respective data products. A common infrastructure for sharing cosmological simulations will reduce the overall number of simulations that need to be produced, reducing the pressure on both the supercomputing facilities and the simulators. It will also precipitate coordination and agreements over who is producing what simulation products and how those products will be utilized and acknowledged within each survey. If the infrastructure also includes a common approach for curating the data and some facilities for analysis, the ability for users to interact with the simulation data directly (rather than through the simulator) will be greatly increased.  

In order to realize any of the common approaches outlined in this report, and to ensure the scientific success of Rubin, WFIRST, and Euclid, it is clear that a common infrastructure needs to be available. This includes hardware (e.g. storage space, data servers, fast connection and transfer links), as well as a common approach for data curation to make data products easily accessible to the community. It also includes expert support personnel (both for the simulations and the data hosting) who are actively engaged in developing and maintaining the infrastructure, in addition to supporting the users. 

\subsection{Examples of Existing Infrastructures}
There are many solutions to hosting and sharing big datasets. The simplest solution is a basic repository that stores and hosts the simulations and associated data products for download by a user. More sophisticated solutions involve utilizing a common approach for data curation and also providing some on-site computing resources to undertake increasingly sophisticated analyses on the data. 

This section provides two examples of existing data sharing and analysis infrastructures that have been used for cosmological simulations. These examples are intended to give some insight into the different solutions available for storing and hosting cosmological simulation data and do not necessarily represent the best solutions for a common Rubin, WFIRST, and Euclid infrastructure. A more detailed and thorough investigation is required to flesh out infrastructure solutions that are optimized for Rubin, WFIRST, and Euclid. 

\subsubsection{Port d'Informaci\'{o} Cient\'{i}fica (PIC)}
The big data system at PIC was used to generate the Euclid
Flagship mock galaxy catalog and to allow collaborative access to this dataset. This full-sky mock, which extends to redshift 2.3, is a dataset of 10s of terabytes. It is made available within the Euclid collaboration via the COSMOHUB\footnote{COSMOHUB https://cosmohub.pic.es/} web portal which allows users to make plots and extract subsets of the data without any prior SQL knowledge. The data processing at PIC is based on the Apache Hadoop file system (HDFS), where data is distributed on local disks of the processing nodes of a compute cluster. This gives very high data rates as long as the I/O processing is always performed on the node that actually contains the relevant data on its local hard disk. By using Apache SPARC, which is a Python based implementation of the map-reduce data processing approach, a very high degree of parallel I/O is sustained across these local hard disks. Generation of the full-sky mock galaxy catalog from a catalog of 40 billion dark matter halos can be achieved in under 24 hours. The software pipeline to produce this mock is called SciPIC, which is written in python using the Apache SPARC framework. One key component of this pipeline is an implementation of the `treecorr' \citep{Jarvis2004} algorithm for estimating galaxy pair correlation functions. This allows the clustering of galaxies in the mock to be calibrated against observational data as a function of luminosity and color. 

While the system is currently able to handle the galaxy and dark matter halo catalogs and the intensity of queries currently coming from within the Euclid consortium, it does not seem suited to handling the raw simulation data. For example, the task of producing the input dark matter halo catalog is done as a separate step at the University of Zurich using a pipeline of specialized parallel codes that can deal with 250 TB of raw particle data. Rewriting halo finders (and other analysis tasks) using Apache SPARC would in principle allow raw data to be handled by a larger system of this type. However, the current splitting of tasks at the dark matter halo catalog level seems to be very efficient, but incurs a lack of transparency and redundant data exchange within the overall process of mock generation.

\subsubsection{The Data-Scope}
In 2010 the NSF awarded a Major Research Instrumentation (MRI) grant to PI Alex Szalay at Johns Hopkins University (JHU) for a project to develop a multi-petabyte generic data analysis environment. This system is called the Data-Scope and it was designed to enable analysis of pertabyte-scale datasets\footnote{The Data-Scope http://idies.jhu.edu/resources/datascope/}. The system has a ~5 petabyte storage capacity and a sequential I/O\footnote{Sequential I/O means that the data must be accessed in order, from the start of the file to the end, while random I/O allows reading or writing any part of the file at any time.} bandwidth of ~500 gigabytes/second. Each individual project is provided with its own node, enabling the data to be stored in a way that is optimized for that project. One example is storing the data in a SQL server, which is a database solution that enables sequential I/O. The data-scope system can reach ~600 teraflops with its GPUs, which is a key component of this enabling technology that requires new software to be written to undertake the more traditional CPU analyses. For example, undertaking a standard correlation function calculation on galaxy pairs can become prohibitive on a CPU when the number of galaxy pairs becomes very large. A massive 400 trillion galaxy pair correlation function calculation was undertaken on Sloan Digital Sky Survey data that was hosted on a system with very similar facilities to the Data-Scope \citep{TNB+11}. The authors reported that the calculation was hundreds of times faster than the same calculation on traditional CPUs. 

The Data-Scope is an example of how very large datasets, like cosmological simulations, can be curated for intensive analyses and the type of hardware, software, and expertise required to undertake these efforts. However, the system is primarily focused on big data analysis and does not address curating the data for long-term hosting and wide-spread community access. The Data-Scope is still operational and accepts proposals from the community to undertake computationally intensive projects on large datasets.

\subsection{Key Challenges}
There are a number of challenges to developing a common infrastructure. There needs to be a plan for where the simulations are being run with some guarantees that those resources will be available for these efforts. Once the simulations have been completed, an initial analysis may be completed at a different facility, so rapid transfer capabilities of very large datasets need to be in place. Decisions need to be made about what data products are being stored and hosted and how those products are being curated to enable widespread use (i.e. does the data need to be stored sequentially or in a format that enables rapid ingestion by a database?). There are a range of solutions, from simply storing and hosting the flat files for direct download by scientists to analyze on a system that they identify themselves, to more sophisticated database solutions that include access to increasingly powerful analysis hardware at the data center.

\subsection{TACS Findings for Common Infrastructure}
Every section of this report required either the generation or utilization of cosmological simulations to ensure the scientific success of Rubin, WFIRST, and Euclid. With limited resources and expertise available for each of the surveys, coordination between the surveys on which cosmological simulations to produce and a common infrastructure to share the data will clearly contribute to the scientific success of each of the surveys. This approach will also save money in the long-term by reducing the overall number of required simulations and facilitating a common data curation approach that will increase user efficiency in accessing and utilizing the simulations. In order to facilitate effective sharing and utilization of the data products, a central, common, data sharing infrastructure is required. In the absence of such coordination and infrastructure, the onus returns to the individual surveys to produce, analyze, store, and host all of their required simulations, resulting in a much higher demand on already limited computational resources and similar simulations being produced up to three times. 

The work to flesh out the range of solutions for an Rubin, WFIRST, Euclid cosmological simulation data sharing infrastructure requires additional effort. This effort includes scoping and costing the hardware requirements, coordinating with the scientists to identify which data products should be stored and the best methods for curating the data, exploring the methods for accessing the data and options for interfacing with the data, scoping a range of support levels that a host data center could provide and costing those options, and providing detailed proposals that show what capabilities and scientific return can be expected with specific levels of investment. TACS finds that a study should be undertaken in collaboration with data centers to investigate and test solutions for a long-term archival infrastructure for simulated cosmological data products. 

\subsection{Conclusions}
\begin{itemize}
\item \textbf{Phase 1: Scoping of Requirements and Architecture; $\sim$6 months}\\
-- Conduct an assessment of possible shared infrastructure and data curation solutions and identify the best choice. \\
-- Develop requirements for an infrastructure with increasing capabilities (cost points), clearly identifying the increased capabilities at each point and highlighting the optimal choice for the surveys, Agencies, and broader community.\\
-- Present a detailed proposal for a test-bed infrastructure that includes requirements for hardware, data curation, and personnel. Outline the tests that will be undertaken and the metrics that will be used to determine the overall success of the test-bed infrastructure. Provide a rough roadmap for moving from the test-bed to a fully realized infrastructure.

\item \textbf{Phase 2: Building and Exercising a Test-Bed Infrastructure; $\sim$18-24 months}\\
-- Acquire and install required new hardware (including computing, storage, and interconnect capabilities).\\
-- Implement and test the chosen data curation solution.\\
-- Test the data sharing solution.\\
-- If implemented, test the on-site analysis capabilities.\\
-- Present an analysis of the success of the test-bed infrastructure and provide a detailed proposal for a fully realized infrastructure.

\item \textbf{Phase 3: Realizing the Full Data Sharing Infrastructure; $\sim$2022-TBD}\\
During Phase 3, a fully realized archival infrastructure would be deployed. The details on the hardware and personnel required to implement this phase 
will depend critically on what is learned during Phases 1 and 2 of this program. 

\end{itemize}

\vspace{0.4cm}
{\Large{\bf Acknowledgements}}\\
AK was supported by JPL, which is run under contract by the California Institute of Technology for NASA. AK was also supported in part by NASA ROSES grant 12-EUCLID12-0004 and NASA grant 15-WFIRST15-0008. Argonne National Laboratory's work by KH and AH was supported under the U.S. Department of Energy contract DE-AC02-76SF00515.
\bibliographystyle{plainnat}
\bibliography{tacs}

\end{document}